%% file: qrhov10.tex
\documentclass[prd,amsmath,showpacs,preprintnumbers,
               floatfix,twocolumn,letterpaper,superscriptaddress,h-physrev4-1]{revtex4-1}

\usepackage{graphicx}
\usepackage{slashed}
\usepackage{amssymb}
\include{nc}

\begin{document}

\title{Chiral extrapolation beyond the power-counting regime}

\author{J. M. M. Hall} 
\affiliation{Special Research Centre for the Subatomic Structure of Matter 
  (CSSM), School of Chemistry and Physics, University of Adelaide 5005,
  Australia}

\author{F. X. Lee}
\affiliation{Physics Department, The George Washington University, Washington, 
DC 20052, USA}

\author{D. B. Leinweber}
\affiliation{Special Research Centre for the Subatomic Structure of Matter 
  (CSSM), School of Chemistry and Physics, University of Adelaide 5005,
  Australia}

\author{K. F. Liu}
\affiliation{Department of Physics and Astronomy, University of Kentucky, 
Lexington, KY 40506, USA}

\author{N. Mathur}
\affiliation{Department of Theoretical Physics, Tata Institute of 
Fundamental Research, Mumbai, India}


\author{R. D. Young} 
\affiliation{Special Research Centre for the Subatomic Structure of Matter 
  (CSSM), School of Chemistry and Physics, University of Adelaide 5005,
  Australia}
\affiliation{ARC Centre of Excellence for Particle Physics at the Terascale, 
School of Chemistry and Physics, University of Adelaide 5005,
  Australia}

\author{J. B. Zhang}
\affiliation{ZIMP and Department of Physics, Zhejiang University,
   Hangzhou, 310027, P. R. China}

\preprint{ADP-11-5/T727, TIFR/TH/11-01}

\begin{abstract}

Chiral effective field theory 
can provide valuable insight into
 the chiral physics of hadrons when used in conjunction with non-perturbative
 schemes such as lattice 
quantum chromodynamics (QCD).
In this discourse, the attention is focused on extrapolating the mass of the 
$\rho$ meson to the physical pion mass in quenched QCD (QQCD).
With the absence of a known experimental value, 
this serves to demonstrate the ability of the extrapolation scheme
 to make predictions without prior bias.
By using extended effective field theory
  developed previously, 
 an extrapolation is performed using quenched lattice QCD data that 
extends outside the chiral power-counting regime (PCR). 
The method involves an analysis of the renormalization flow curves 
of the low-energy coefficients in a finite-range regularized effective 
field theory. The analysis identifies an optimal regularization scale, which 
is embedded in the lattice QCD data themselves. This optimal scale  
is the value of the regularization scale 
at which the renormalization of the low-energy 
coefficients is approximately independent of the range of quark masses 
considered.
By using recent precision, quenched lattice results, 
the extrapolation is tested directly by truncating the 
 analysis to a set of points above 
 $380$ MeV, while temporarily disregarding the simulation results closer to the 
 chiral regime. This tests the ability of the method to make 
 predictions of the simulation results, without phenomenologically 
 motivated bias. 
 The result is a successful extrapolation to the chiral regime.


\end{abstract}

\maketitle

\section{Introduction}
\label{sect:intro}

In lattice quantum chromodynamics (QCD), the calculation of
observables with light dynamical quarks is computationally intensive,
and only in recent times have there been successful attempts to
perform calculations of any observable at the physical point ($m_\pi =
140$ MeV) \cite{Aoki:2008sm,Kuramashi:2008tb,Durr:2010aw}.  
Usually, some extrapolation scheme is
needed if one is to compare theoretical calculations with the
corresponding physical observables. Utilizing lattice QCD results spread  
over a larger range of quark masses naturally enables greater statistical 
precision in the extrapolation.

Quenched QCD (QQCD) was introduced as a way to ameliorate the computational
difficulty of simulating dynamical fermions on the lattice.
Quenched simulations typically have been superseded by the wide 
availability of dynamical configurations. Nevertheless, it can 
still be used as an efficient testing ground.
This is particularly true of the chiral extrapolation problem, where 
the experimentally known values may introduce a prejudice on a chosen 
form. In QQCD, the physical target point does not exist, and an extrapolation 
of moderate-mass points to the chiral regime provides an unbiased test 
of the procedure.

In order to discuss the chiral behaviour of the $\rho$ meson in QQCD, one
first constructs an effective field theory describing the relevant low-energy 
degrees of freedom.  The mass of the $\rho$ meson is described
by a chiral expansion in the quark mass ($m_q$), which includes
analytic terms that are polynomial in $m_q$, and non-analytic terms
arising from chiral loop integrals.  These loop integrals are commonly
divergent, and thus it is necessary to introduce a regularization
procedure.
Finite-range regularization (FRR) is selected as a regularization
scheme, which introduces a momentum cutoff scale $\La$ into the loop
integrals. The properties of FRR allow it to be used with data
extending outside the power-counting regime (PCR),
 at the expense of complete scheme-independence.
  As has been demonstrated, an optimal choice of
regularization scale, $\La_{\ro{scale}}$, 
can be extracted from the lattice simulation results
\cite{Hall:2010ai}. A systematic uncertainty in $\La_{\ro{scale}}$ can
also be estimated, which provides a range of suitable values for the
scale obtained from the data \cite{Young:2009ub}.  Thus the scheme-dependence 
in using data extending outside the PCR can be quantified
in an unbiased fashion.
 

\section{Extended effective field theory}
\label{sect:eft}

\begin{figure}[tp]
\centering
\includegraphics[width=0.55\hsize]{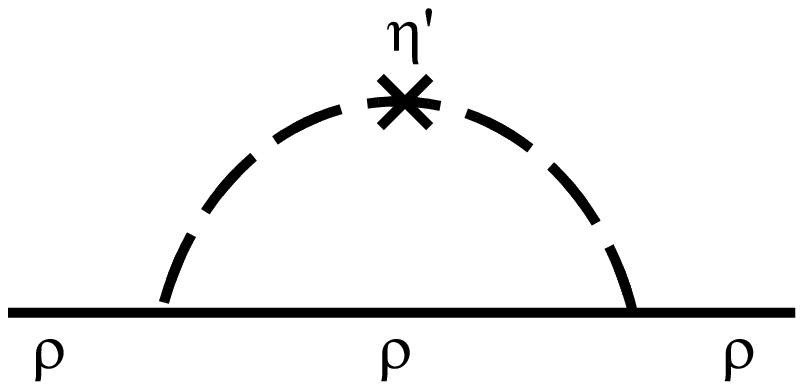}
\vspace{-6pt}
\caption{\footnotesize{Double hairpin $\eta'$ diagram.}}
\label{fig:qrhoSEdh}
\vspace{10pt}
\centering
\includegraphics[width=0.55\hsize]{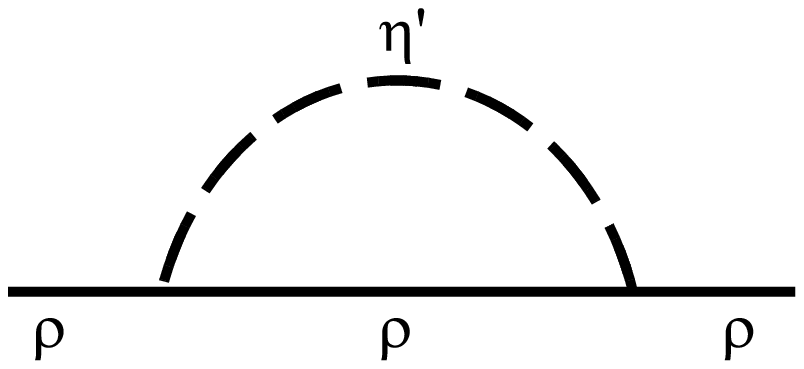}
\vspace{-6pt}
\caption{\footnotesize{Single hairpin $\eta'$ diagram.}}
\label{fig:qrhoSEsh}
\end{figure}

\begin{figure}[tp]
\centering
\includegraphics[width=0.55\hsize]{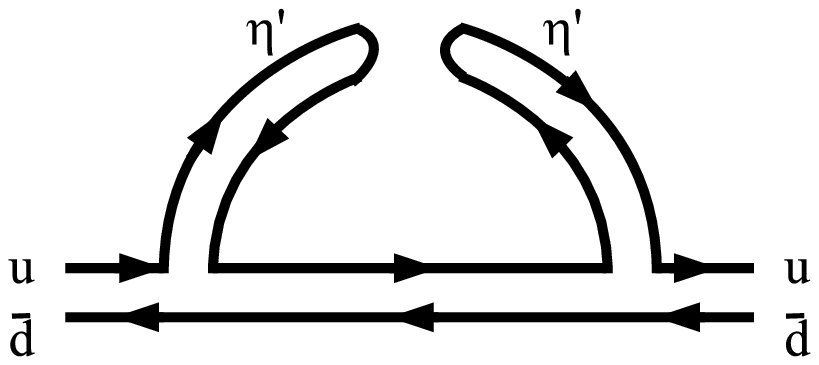}
\vspace{-6pt}
\caption{\footnotesize{Double hairpin quark flow diagram.}}
\label{fig:qrhoQFdh}
\vspace{10pt}
\centering
\includegraphics[width=0.55\hsize]{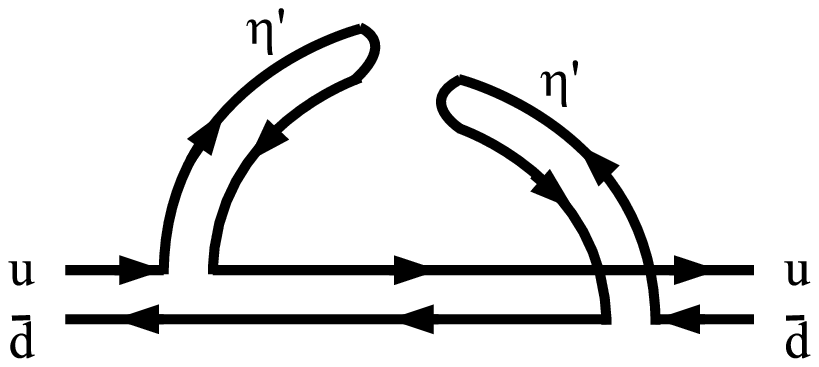}
\vspace{-6pt}
\caption{\footnotesize{Alternative double hairpin quark flow diagram.}}
\label{fig:qrhoQFdhalt}
\vspace{10pt}
\centering
\includegraphics[width=0.55\hsize]{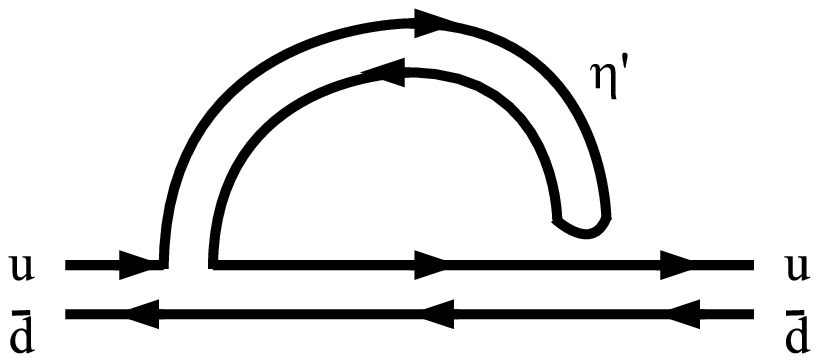}
\vspace{-6pt}
\caption{\footnotesize{Single hairpin quark flow diagram.}}
\label{fig:qrhoQFsh}
\end{figure}

In chiral effective field theory ($\chi$EFT), 
the diagrammatic formulation can be used to identify the
major contributions to the $\rho$ meson mass in QQCD
\cite{Chow:1997dw,Armour:2005mk}.  The leading-order diagrams are the
double and single $\eta'$ hairpin diagrams as shown in Figures
\ref{fig:qrhoSEdh} and \ref{fig:qrhoSEsh}, respectively.
The constant coefficients of these loop integrals are endowed with 
an uncertainty to encompass the possible effects of smaller contributions 
to order $\ca{O}(m_\pi^4)$. 
Interactions with the flavour-singlet $\eta'$ are  
the most important contributions to the $\rho$ meson mass in QQCD.
This is an artifact of the quenched approximation, where the $\eta'$
also behaves as a pseudo-Goldstone boson, having a ``mass'' that is
degenerate with the pion.
The dressing of the $\rho$ meson by the $\eta'$ field is illustrated in
Figures \ref{fig:qrhoQFdh} through \ref{fig:qrhoQFsh}.  
 Since the hairpin vertex must be a flavor-singlet, 
the 
 mesons that can contribute are 
the $\eta'$ meson, and the $\omega$ meson. 
The contributions from the $\omega$ meson are insignificant 
due to OZI suppression and the small $\rho$-$\omega$ mass splitting. 
However, in QQCD, the $\eta'$ loop behaves much as a pion loop,
yet with a slightly modified propagator.

\begin{figure}[tp]
\begin{center}
\includegraphics[width = 0.65\hsize]{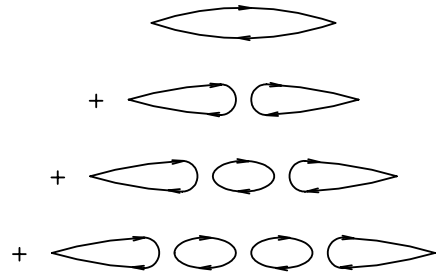}
\vspace{-6pt}
\caption{\footnotesize{Diagrammatic representation of $\eta'$ 
propagator terms.}}
\label{fig:PQetaPrime}
\end{center}
\end{figure}

In full QCD however, the $\eta'$ does not play any role in the
low-energy dynamics.  The physical $\eta'$ acquires a finite mass ---
which survives in the chiral limit --- by re-summing the chain of
vacuum insertions as depicted in Figure~\ref{fig:PQetaPrime}.
As a ``heavy'' degree of freedom, the $\eta'$ can then be integrated
out of the of the effective field theory.
%
%
%

\subsection{Loop integrals and definitions}
\label{subsect:loop}

Using the Gell-Mann$-$Oakes$-$Renner
Relation connecting quark and pion masses (assuming negligible anomalous 
scaling),
$m_q \propto m_\pi^2$ \cite{GellMann:1968rz}, 
the $\rho$ meson mass extrapolation formula in QQCD can be expressed
in a form that contains an analytic polynomial in $m_\pi^2$ plus 
the chiral loop integrals ($\Si^Q$):
\begin{align}
m_{\rho,Q}^2 &= a_0 + a_2m_\pi^2 +a_4m_\pi^4\nn\\
 &+ \Si_{\eta'\!\eta'}^Q(m_\pi^2,\La)
 + \Si_{\eta'}^Q(m_\pi^2,\La) +\ca{O}(m_\pi^5).
\label{eqn:mrhoexpsn}
\end{align}
The coefficients $a_i$ are the `residual series' coefficients, which
correspond to direct quark-mass insertions in the underlying
Lagrangian of chiral perturbation theory. 
 However, the non-analytic behaviour of the expansion arises from the 
chiral loop 
integrals. 
Upon renormalization of the divergent loop
integrals, these will correspond with low-energy constants of the
quenched $\chi$EFT. 
The extraction of these parameters 
from lattice QCD results will now be demonstrated.

By convention, the non-analytic terms from the double and single
hairpin integrals are $\chi_1 m_\pi$ and $\chi_3 m_\pi^3$,
respectively.  The coefficients $\chi_1$ and $\chi_3$ of the leading-order 
non-analytic terms 
are scheme-independent constants that can be 
estimated from phenomenology. 
The low-order expansion of the loop contributions takes the following form:
\begin{align}
\label{eqn:doubloopexpsn}
\Si_{\eta'\!\eta'}^Q &= b^{\eta'\!\eta'}_0 + \chi_1 m_\pi + 
b^{\eta'\!\eta'}_2 m_\pi^2 + \chi_3^{\eta'\!\eta'} m_\pi^3 
+ b^{\eta'\!\eta'}_4 m_\pi^4 \nn\\
&+ \ca{O}(m_\pi^6),\\
\Si_{\eta'}^Q &= b^{\eta'}_0 + b^{\eta'}_2 m_\pi^2 + \chi_3^{\eta'} m_\pi^3
 + b^{\eta'}_4 m_\pi^4 + \ca{O}(m_\pi^6),
\label{eqn:singloopexpsn}
\end{align}
The coefficient $\chi_3$ is obtained by adding the contributions from 
both integrals, 
 $\chi_3 = \chi_3^{\eta'\!\eta'} + \chi_3^{\eta'}$. 
Each integral has a solution in the form of a polynomial expansion
analytic in $m_\pi^2$ plus non-analytic terms, of which the leading-order 
term is of greatest interest.  The coefficients
$b_i$ are scale-dependent and therefore scheme-dependent. 
In order to achieve an extrapolation
based on an optimal FRR scale, first the scale-dependence of the
low-energy expansion must be removed through renormalization.  The
renormalization program of FRR combines the scheme-dependent $b_i$
coefficients from the chiral loops with the scheme-dependent $a_i$
coefficients from the residual series at each chiral order $i$.  The
result is a scheme-independent coefficient $c_i$:
\begin{align}
c_0 &= a_0 + b_0^{\eta'\!\eta'} + b_0^{\eta'},\\
c_2 &= a_2 + b_2^{\eta'\!\eta'} + b_2^{\eta'},\\
c_4 &= a_4 + b_4^{\eta'\!\eta'} + b_4^{\eta'}, \mbox{\,\,etc.}
\end{align}
That is, the underlying $a_i$ coefficients undergo a renormalization
from the chiral loop integrals. The renormalized coefficients $c_i$
are an important part of the extrapolation technique. A stable and
robust determination of these parameters
forms the core of determining an optimal scale $\La^\ro{scale}$.

The loop integrals can be expressed in a convenient form by taking the
non-relativistic limit and performing the pole integration for $k_0$.
Renormalization is achieved by subtracting the relevant terms in the
Taylor expansion of the loop integrals and absorbing them into the
corresponding low-energy coefficients, $c_i$:
\begin{align}
\label{eqn:doub}
\tilde{\Si}_{\eta'\!\eta'}^Q(m_\pi^2;\La) &= \f{-\chi_{\eta'\!\eta'}}
{3\pi^2}\int\!\!\ud^3
 k
\f{(M_0^2 k^2 +\f{5}{2}A_0 k^4)u^2(k;\La)}{{(k^2 + m_\pi^2)}^2} \nn\\
&- b_0^{\eta'\!\eta'}
 - b_2^{\eta'\!\eta'}m_\pi^2
-b_4^{\eta'\!\eta'}m_\pi^4,\\
\label{eqn:sing}
\tilde{\Si}_{\eta'}^Q(m_\pi^2;\La) &= \f{\chi_{\eta'}}{2\pi^2}\int\!\!\ud^3 k
\f{k^2 u^2(k;\La)}{k^2 + m_\pi^2} - b_0^{\eta'} - b_2^{\eta'}m_\pi^2\nn\\
&-b_4^{\eta'}m_\pi^4.
\end{align}
The tilde ($\,\tilde{\,}\,$) denotes that the integrals are written
out in renormalized form to chiral order $\ca{O}(m_\pi^4)$.  
The coefficients $\chi_{\eta'\!\eta'}$ and $\chi_{\eta'}$ are related to the 
 coefficients of the leading-order non-analytic terms by:
\begin{align}
\chi_1 &= M_0^2\,\chi_{\eta'\!\eta'},\\
\chi_3 &= \chi_3^{\eta'\!\eta'} + \chi_3^{\eta'}  
= A_0\,\chi_{\eta'\!\eta'} + \chi_{\eta'}.
\end{align}
%
These couplings are discussed in detail below. 
The
function $u(k;\La)$ is a finite-range regulator with cutoff scale
$\La$, which must be normalized to $1$ at $k^2=0$, and must approach $0$
sufficiently fast to ensure convergence of the loop.  Different
functional forms of $u(k;\La)$ are equivalent within the PCR
\cite{Young:2002ib,Leinweber:2005xz}.  Different choices of $u(k;\La)$
for this investigation are discussed in Sec.~\ref{subsect:frr}.

With the loop integrals specified, Eq.~(\ref{eqn:mrhoexpsn}) can be
rewritten in terms of the renormalized coefficients $c_i$:
\begin{align}
\label{eqn:renormexpsn1}
m_{\rho,Q}^2 &= c_0 + c_2m_\pi^2 + c_4m_\pi^4 
+ \tilde{\Si}_{\eta'\!\eta'}^Q(m_\pi^2;\La) \nn\\
&+ \tilde{\Si}_{\!\eta'}^Q(m_\pi^2;\La) 
+ \ca{O}(m_\pi^5)\\
&\approx
c_0 + \chi_1 m_\pi + c_2 m_\pi^2 + \chi_3 m_\pi^3 + c_4 m_\pi^4 \nn\\
&+ \ca{O}(m_\pi^5).
\label{eqn:renormexpsn2}
\end{align}
Eq.~(\ref{eqn:renormexpsn1}) will be used as 
 the extrapolation formula for $m_{\rho,Q}$ 
at infinite lattice volume. 
The fit coefficients are $c_0$, $c_2$ and $c_4$, 
and $m_{\rho,Q}$ is obtained by taking the square root of 
Eqs.(\ref{eqn:renormexpsn1}) and (\ref{eqn:renormexpsn2}).
It is important to note that the formula in Eq.~(\ref{eqn:renormexpsn2}) 
 is equivalent to Eq.~(\ref{eqn:renormexpsn1}) only as $\La$ is taken to 
infinity. 


Since lattice simulations are necessarily carried out on a discrete
spacetime, any extrapolations performed should take into account
finite-volume effects.  The low-energy effective field theory 
is ideally suited for
characterising the leading infrared effects associated with the finite
volume. In order to achieve this, each of the three-dimensional integrals
can be transformed to its form on the lattice using a finite sum of
discretized momenta, following Armour \textit{et al.}  \cite{Armour:2005mk}, 
for instance:
\eqb
\int\!\!\ud^3 k \ra \f{{(2\pi)}^3}{L_x L_y L_z} \sum_{k_x,k_y,k_z}. 
\eqe
Each momentum component is quantized in units of $2\pi/L$, that
is $k_i=n_i2\pi/L$ for integers $n_i$.  Finite-volume corrections 
$\de^{\ro{FVC}}$
 can be written simply as the difference between the finite sum
and the corresponding integral. It is known that the finite-volume 
corrections saturate to a fixed result for large values of the 
regularization scale  
\cite{Hall:2010ai}.
Following the example set by this article, the value $\La'= 2.0$ GeV is chosen
to evaluate all finite-volume corrections independent of the  
FRR cutoff
 scale $\La$ in Eqs.(\ref{eqn:doub}) and (\ref{eqn:sing}).
The finite-volume version of Eq.~(\ref{eqn:renormexpsn1})
 can thus be expressed:
\begin{align}
m_{\rho,Q}^2 &= c_0 + c_2m_\pi^2 + c_4m_\pi^4 
+ (\tilde{\Si}_{\eta'\!\eta'}^Q (m_\pi^2;\La)\nn\\
&+ \de^\ro{FVC}_{\eta'\!\eta'}(m_\pi^2;\La')) 
+ (\tilde{\Si}_{\!\eta'}^Q (m_\pi^2;\La) + \de^\ro{FVC}_{\eta'}(m_\pi^2;\La'))\nn\\
&+ \ca{O}(m_\pi^5).
\label{eqn:renormexpsnfin}
\end{align}
%

\begin{figure}[tp]
\begin{center}
\includegraphics[width = 0.80\hsize]{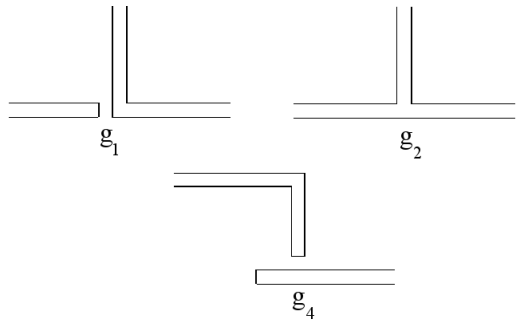}
\vspace{-6pt}
\caption{\footnotesize{Coupling types following convention 
introduced by Booth \cite{Booth:1996hk}.}}
\label{fig:coupl}
\end{center}
\end{figure}

The convention used for defining the values of $\chi_1$, $\chi_3$, and
the various coupling constants that occur in each, 
follows Booth \cite{Booth:1996hk}. For the possible different
values that coupling constants can take, 
definitions by Chow \& Rey \cite{Chow:1997dw},
 Armour \textit{et al.} \cite{Armour:2005mk} and
Sharpe \cite{Sharpe:1996ih} are used.
The types of vertices available are shown in Figure \ref{fig:coupl},
where $g_2$ and $g_4$ occur explicitly in the two diagrams considered here.
Booth suggests naturalness for  $g_2 \sim 1$, and that $g_4 \sim 1/N_c$.
These quenched coupling constants can be connected with the experimental
value of $g_{\om\rho\pi}$ as per Lublinsky \cite{Lublinsky:1996yf} by
the relation:
\eqb
g_2 = \f{1}{2}g_{\om\rho\pi}f_\pi,
\eqe
where $g_{\om\rho\pi} = 14\pm2$ GeV$^{-1}$ and
the pion decay constant takes the value $f_\pi = 0.0924$ GeV.
Thus $g_2$ is chosen to be $0.65\pm0.09$ GeV and $g_4$ is chosen to be 
approximately $g_2/3$.
The coupling between the separate legs of the double hairpin
diagram are approximated by the massive constant $M_0^2 \propto m_{\eta'}^2$.
The next-order correction to $M_0$ in momentum $k$ defines the 
coupling to be $-M_0^2 + A_0 k^2$. These constants can be connected to
the full QCD $\eta'$ meson mass $m_{\eta'}$ by considering the geometric
series of terms as previously illustrated in Figure \ref{fig:PQetaPrime}.
%
%
For the value of $M_0$, Booth suggest $M_0 \approx 400$ MeV  
by comparing the estimate from a hairpin insertion to the result from the  
Witten-Veneziano formula \cite{Booth:1996hk}. In a paper by 
Duncan \textit{et al.} a value of $M_0 \approx 900$ MeV is obtained 
if the coupling constant $A_0$ is natural. Furthermore, an analysis 
of the topological susceptibility leads to an estimate 
$M_0 = 1.1 \pm 0.2$ GeV \cite{Duncan:1996ma}. 
In this analysis, an average value $M_0 = (400+900+1100)/3 = 800$ MeV 
is sensible as a first approximation. 
As a further check, consider 
the formula from Ref.~\cite{Duncan:1996ma}, using our 
normalization for the pion decay constant ($f_\pi^2 = 2 f_{\pi,\ro{Duncan}}^2$):
\eqb
\label{eqn:M0}
\delta = \f{A_0 M_0^2}{48 \pi^2 f_\pi^2}.
\eqe
This formula relates the couplings $A_0$ and $M_0^2$ to the anomalous scaling 
parameter of the pion mass in quenched QCD, defined by:
\eqb
m_\pi^2 \approx m_q^{\f{1}{1+\delta}}.
\eqe
The parameter $\delta$ is found to be 
 small (and the Gell-Mann$-$Oakes$-$Renner
Relation a good approximation),
with a maximum value estimated by Duncan to be 
$\delta_{\ro{max}} = 0.03$ \cite{Duncan:1996ma}. 
%
Booth comments that the parameter $A_0$ is small, and 
 vanishes in the limit 
$N_c \rightarrow \infty$. Nevertheless, Sharpe uses a finite value 
$A_0 \sim 0.2$ \cite{Sharpe:1996ih}. 
By using these finite values for $\delta$ and $A_0$, Eq.~(\ref{eqn:M0})
leads to a value of $M_0^2\approx 0.6$ GeV$^2$. 
As a result, $M_0^2$ is taken to be $0.6\pm0.2$ GeV$^2$ and 
$A_0$ is taken to be $0\pm0.2$.

The coefficients $\chi_{\eta'\!\eta'}$ and $\chi_{\eta'}$ 
can be specified in terms of the
relevant coupling constants:
\begin{align}
\chi_{\eta'\!\eta'} &= -2\mr\!\f{g_2^2}{4\pi f_\pi^2},\nn\\
\chi_{\eta'} &= -2\mr\!\f{g_2 g_4}{6\pi f_\pi^2},
\end{align}
where the couplings are defined relative to 
$\mr$ representing the $\rho$ meson mass in the chiral limit, which 
is taken to be $770$ MeV. 
%

\subsection{Finite-range regularization}
\label{subsect:frr}

In FRR, regulator functions $u(k;\La)$ with characteristic scale $\La$ 
are inserted into the
loop integrals to control the ultraviolet divergences that occur in
the loop integrals encountered. 
%
For some choices of regulator, extra regulator-dependent 
non-analytic terms arise in the chiral expansion of
Eq.~(\ref{eqn:renormexpsn2}).
Since the correct non-analytic terms of the chiral expansion are 
regularization scale-independent terms, 
the extra non-analytic terms within working chiral order must be 
removed. All scale-dependence should be absorbed into the analytic fit 
parameters $a_i$.
%
For example, if a dipole regulator is chosen, the extra terms 
$b_3^{(1)}m_\pi^3$, $(b_5^{(1)} + b_5^{(3)})m_\pi^5$ and higher-order terms 
  occurring at odd powers of $m_\pi$ feature in 
Eq.~(\ref{eqn:renormexpsn2}). 
One can avoid this by choosing a regulator that does not generate these
extra terms, up to working-order $\ca{O}(m_\pi^4)$.
Since the step function
 $u^2(k;\La) = \theta(\La-k)$ introduces inconvenient finite-volume artifacts,
a `triple-dipole' form factor will be chosen, defined by:
%
%
\eqb
u(k;\La) = {\left(1 + {\left[\f{k^2}{\La^2}\right]}^3\right)}^{-2}.
\eqe
%


%

\section{Lattice simulation details}
The calculation is performed on a $20^3 \times 32$ lattice with 197 gauge configurations generated with the Iwasaki gauge action \cite{Iwasaki:1985we} 
 at $\beta = 2.264$, and the quark propagators are calculated with overlap fermions and a wall-source technique. The lattice spacing is  $0.153$ fm, as determined from the Sommer scale parameter.

The massive overlap Dirac operator is defined \cite{Neuberger:1997bg} 
in the following way so that at 
tree-level there is no mass or wavefunction renormalization \cite{Dong:2001fm}: 
\begin{equation} \label{over-op}
D(m) = \rho + \frac{m}{2} + (\rho - \frac{m}{2} ) \gamma_5 \epsilon (H),
\end{equation}
where $\epsilon(H)$ is the matrix sign function of an Hermitian operator $H$.
 $\epsilon(H)\equiv
H_W/|H_W| = H_W/(H_W^\dagger H_W)^{1/2}$ is chosen, where $H_W(x,y)=\gamma_5
D_W(x,y)$. Here $D_W$ is the
usual Wilson-Dirac operator on the lattice, except with a
negative mass parameter $- \rho = 1/2\kappa -4$ in which $\kappa_c < \kappa <
0.25$. Taking $\kappa = 0.19$ in the calculation corresponds to $\rho =
1.368$ \cite{Chen:2003im,Zhang:2005sca}. 

In Figure \ref{fig:data} the simulation results for the vector meson mass 
are shown for a range of quark masses.

\begin{figure}[tp]
\includegraphics[width=0.80\hsize,angle=0]{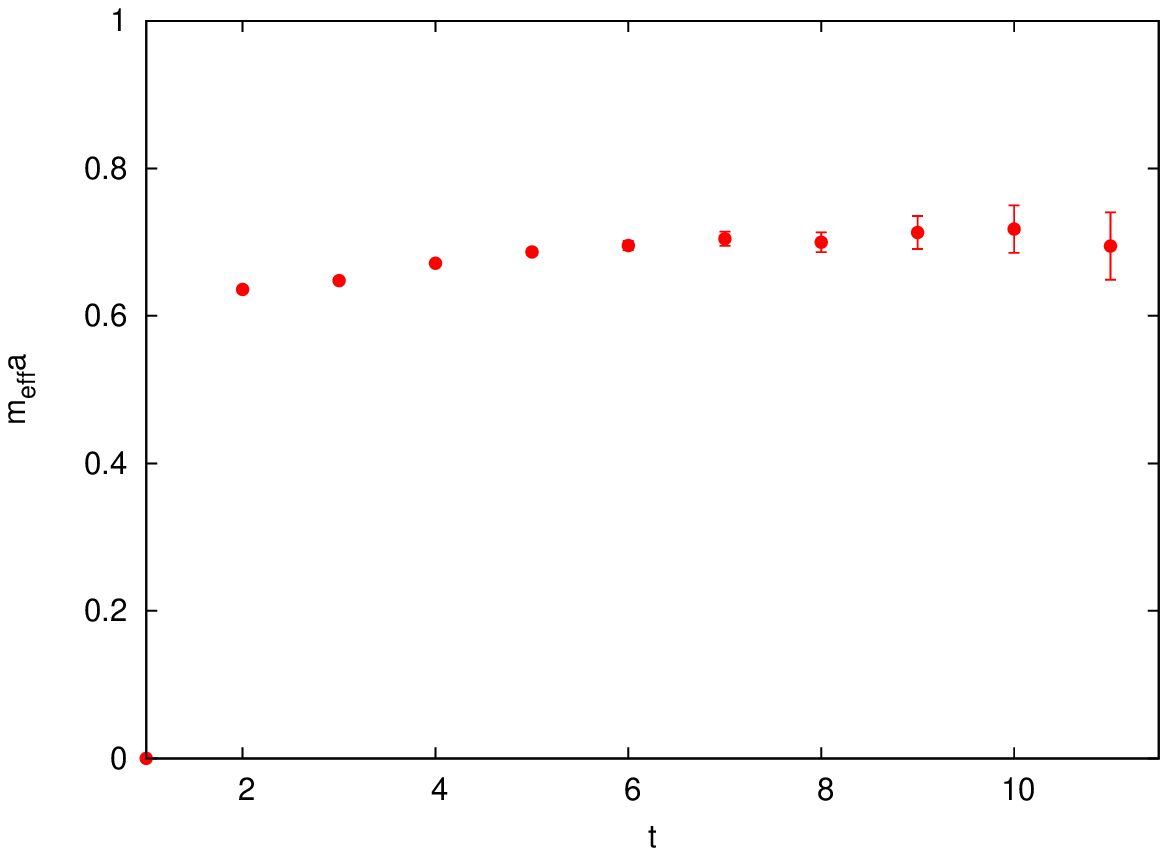}
\vspace{-11pt}
\caption{\footnotesize{(color online). Effective mass plot corresponding to the simulation at $m_\pi^2 = 0.143$ GeV$^2$. Only the wall source results are plotted. The point source results are not used in the analysis.}}
\label{fig:em18}
\includegraphics[width=0.80\hsize,angle=0]{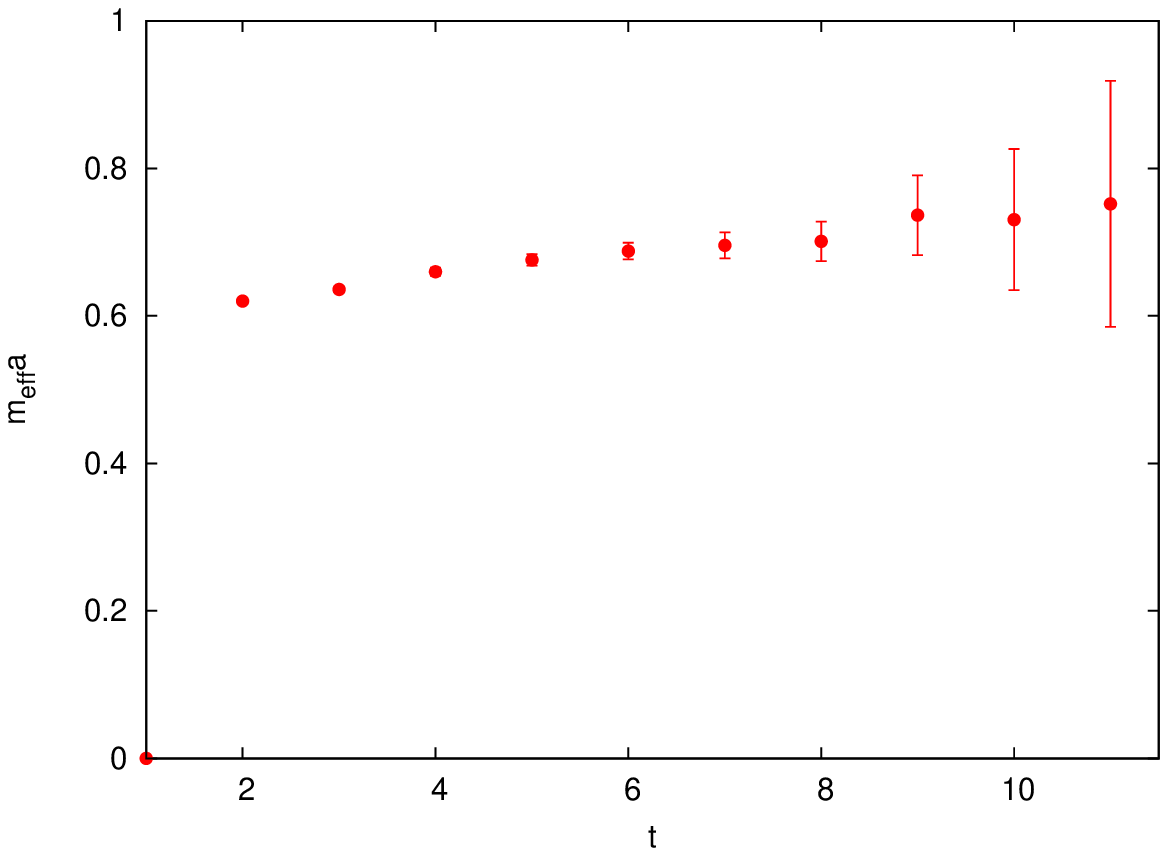}
\vspace{-11pt}
\caption{\footnotesize{(color online). Effective mass plot corresponding to the simulation at $m_\pi^2 = 0.080$ GeV$^2$. Only the wall source results are plotted. The point source results are not used in the analysis.}}
\label{fig:em21}
\includegraphics[width=0.80\hsize,angle=0]{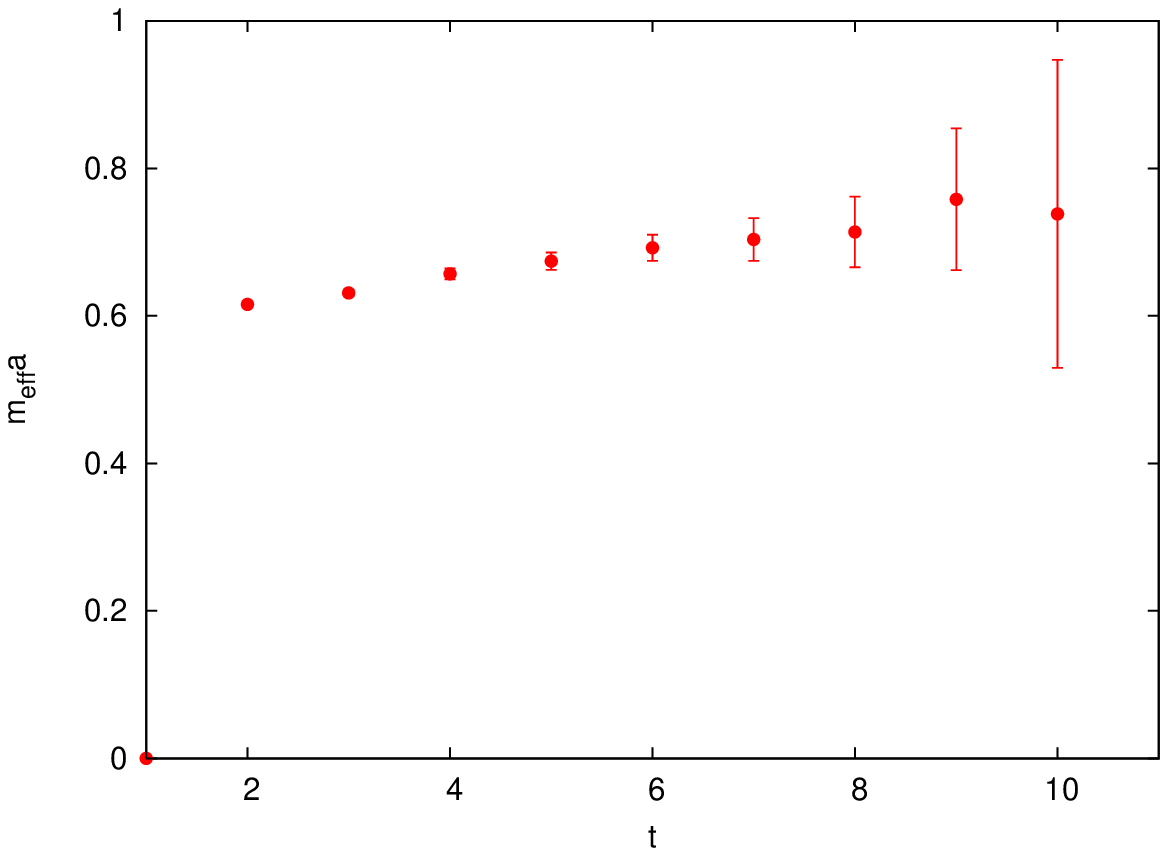}
\vspace{-11pt}
\caption{\footnotesize{(color online). Effective mass plot corresponding to the simulation at $m_\pi^2 = 0.053$ GeV$^2$. Only the wall source results are plotted. The point source results are not used in the analysis.}}
\label{fig:em24}
\includegraphics[width=0.80\hsize,angle=0]{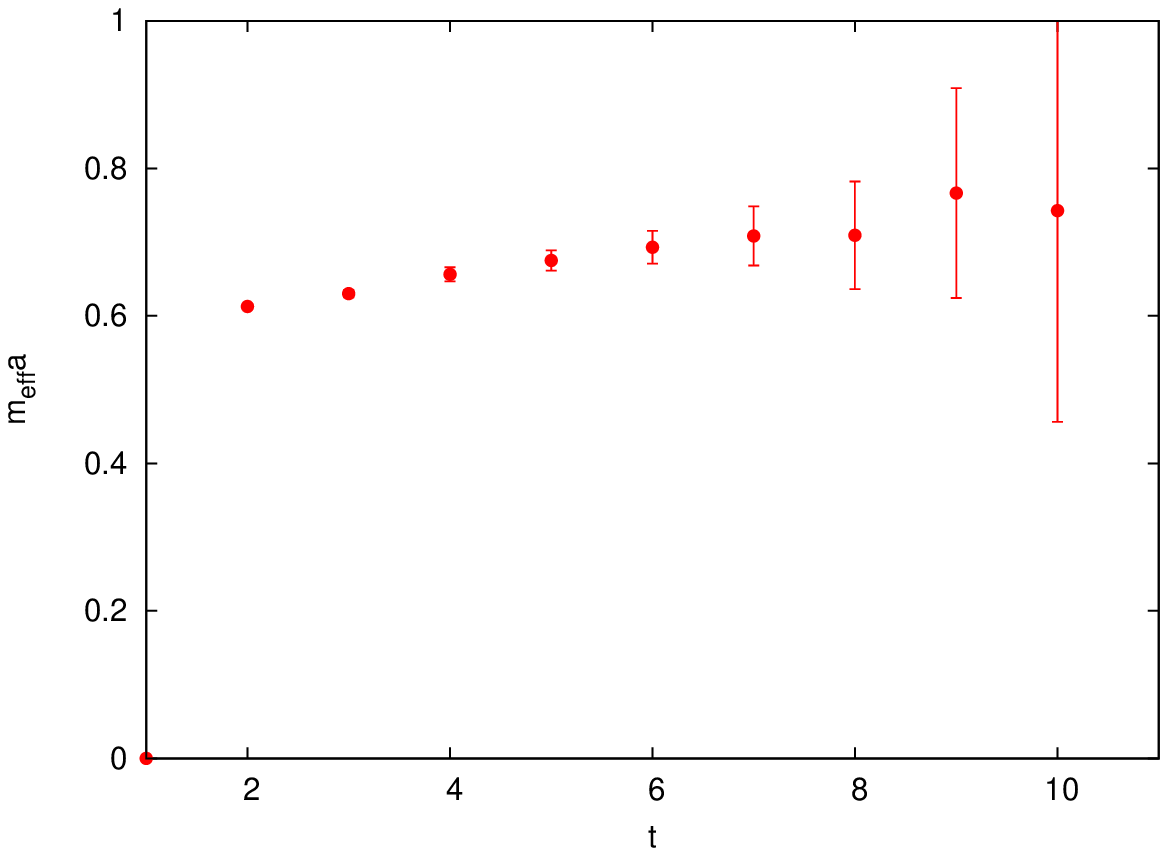}
\vspace{-11pt}
\caption{\footnotesize{(color online). Effective mass plot corresponding to the simulation at $m_\pi^2 = 0.041$ GeV$^2$. Only the wall source results are plotted. The point source results are not used in the analysis.}}
\label{fig:em26}
\end{figure}

The data displayed in Figure \ref{fig:data} 
are split into two parts. All the data 
left of the solid vertical line is unused for extrapolation and 
kept in reserve. 
Indeed, the authors performing the extrapolation were blind to these data. 
This is so that the extrapolation can be checked against 
 these known data points once the extrapolation is established. 
In other words, the results of the chiral extrapolation are genuine 
\textit{predictions} of the hidden lattice results. 
Only 
the data points to the right of the solid vertical 
line are used for extrapolation. 
The full set of data is also listed in Table 
\ref{table:rhodata}, which also includes the bare quark mass values. 
In addition, effective mass plots corresponding to 
four lighter pion masses are included, in Figures \ref{fig:em18} 
through \ref{fig:em26}. 

\begin{table}[tp]
 \caption{\footnotesize{Quenched lattice QCD data for
    the $\rho$ meson mass $m_\rho$
 at various pion mass squared values $m_\pi^2$. 
The statistical uncertainty of the $m_\pi^2$ is negligible. 
The values of the bare quark mass $m_q$ are also included 
for comparison. 
The lattice size is 
 $20^3 \times 32$, with a lattice spacing of $0.153$ fm. 
Entries below the line (underneath $m_\pi^2 = 0.143$ GeV$^2$) 
remained hidden until the extrapolation 
was determined.
}}
  \newcommand\T{\rule{0pt}{2.8ex}}
  \newcommand\B{\rule[-1.4ex]{0pt}{0pt}}
  \begin{center}
    \begin{tabular}{llll}
      \hline
      \hline
       \T\B 
      $m_q$ (GeV) & $m_\pi^2$(GeV$^2$) &  $m_\rho$(GeV)  & \quad$m_\pi L$  \\
      \hline
$1.032$ & $3.150$ & $2.001(1) $ & \quad$27.53$ \\
$0.774$ & $2.187$ & $1.700(2) $ & \quad$22.94$ \\
$0.645$ & $1.742$ & $1.548(2) $ & \quad$20.47$ \\
$0.516$ & $1.329$ & $1.399(2) $ & \quad$17.88$ \\
$0.477$ & $1.212$ & $1.354(2) $ & \quad$17.08$ \\
$0.426$ & $1.062$ & $1.294(2) $ & \quad$15.98$ \\
$0.356$ & $0.867$ & $1.214(3) $ & \quad$14.44$ \\
$0.309$ & $0.743$ & $1.162(4) $ & \quad$13.37$ \\
$0.284$ & $0.676$ & $1.133(4) $ & \quad$12.75$ \\
$0.258$ & $0.610$ & $1.103(5) $ & \quad$12.12$ \\
$0.219$ & $0.515$ & $1.060(5) $ & \quad$11.13$ \\
$0.181$ & $0.422$ & $1.016(6) $ & \quad$10.07$ \\
$0.148$ & $0.347$ & $0.985(7) $ & \quad$9.13$ \\
$0.123$ & $0.288$ & $0.960(8) $ & \quad$8.32$ \\
$0.102$ & $0.241$ & $0.938(8) $ & \quad$7.62$ \\
$0.085$ & $0.204$ & $0.926(9) $ & \quad$7.00$ \\
$0.071$ & $0.172$ & $0.914(11) $ & \quad$6.43$ \\
$0.058$ & $0.143$ & $0.908(14) $ & \quad$5.87$ \\
\hline
$0.045$ & $0.114$ & $0.899(15) $ & \quad$5.24$ \\
$0.036$ & $0.094$ & $0.899(16) $ & \quad$4.75$ \\
$0.030$ & $0.080$ & $0.896(18) $ & \quad$4.38$ \\
$0.025$ & $0.068$ & $0.898(20) $ & \quad$4.04$ \\
$0.021$ & $0.059$ & $0.902(22) $ & \quad$3.77$ \\
$0.018$ & $0.053$ & $0.903(26) $ & \quad$3.58$ \\
$0.015$ & $0.047$ & $0.907(28) $ & \quad$3.37$ \\
$0.013$ & $0.041$ & $0.913(32) $ & \quad$3.15$ \\
      \hline
    \end{tabular}
  \end{center}
\vspace{-6pt}
  \label{table:rhodata}
\end{table}

  To estimate 
finite-volume effects using overlap fermions, quenched lattices 
of volumes $16^3 \times 28$ and $12^3 \times 28$ with $a = 0.2$ fm 
are used. For a pion mass of $180$ MeV, $m_{PS}L \approx 3$, and 
the finite-volume correction is approximately $2.7$ MeV: about $1.5$\% 
of the pion mass \cite{Chen:2003im}.
The current $20^3 \times 32$ lattice with 
$a = 0.153$ fm is about the same physical
size as that of a $16^3 \times 28$ lattice and a similar 
finite-volume correction is expected. 
 To estimate the finite-volume correction of the lowest $\rho$ meson mass at 
$m_\pi \approx 200$ MeV, 
the same percentage
of error is used, and a shift of $\delta_L m_\rho \approx 13$ MeV 
to the $\rho$ mass is calculated 
for the $\rho$ meson mass
of $m_\rho \approx 917$ MeV. 
This is about half of the statistical error of the lattice data. It 
should be noted that
the data that will be used in chiral extrapolations are those with pion mass 
greater than $400$ MeV, with $m_{PS}L > 6.2$. The predictions are 
extended to the
region with pion mass less than $400$ MeV and compared with the lattice data.

With regards to possible lattice artifacts, 
the lattice results analyzed are based on the overlap fermion on
quenched gauge configurations at one lattice spacing. Even though
the overlap fermion has relatively smaller $\ca{O}(a)$ errors, 
the $\ca{O}(a^2)$ correction toward the continuum limit has not been 
taken into account. With a spatial size of $3.06$ fm,  
$m_{\pi} a \sim 3$ for the smallest pion mass
at $200$ MeV is somewhat
smaller than $m_{\pi} a = 4$, 
beyond which the finite volume effect has been considered
to be small. For $m_{\pi}a \sim 3$, the previous study described in 
Ref.~\cite{Chen:2003im} 
estimates that
the finite-volume correction is approximately $3\%$ 
which is smaller than
the statistical error of the pion mass.

The 
enhancement of zero modes effects in QQCD 
primarily affects the pseodoscalar and scalar mesons.
Since all the zero modes appear in one chiral sector in each gauge 
configuration,
the pseodoscalar and scalar mesons will have a leading $1/m^2$ singularity 
from the
zero modes. These appear in both the quark and antiquark propagators in the 
meson
correlator \cite{Dong:2001fm}.
Nevertheless, the vector and axial vector mesons have only a  
$1/m$ singularity, which is a less dramatic effect. In
either case, the quantity that determines the size of the zero mode effects 
 is
$m\Sigma V$ in the $p$-regime \cite{Leutwyler:1992yt}. 
It has been demonstrated that when $m\Sigma V \gtrsim 5$, the zero mode 
effect is hardly
detectable \cite{Chen:2003im,Li:2010pw}. 
For all pion masses displayed in Figure \ref{fig:data}, 
$m\Sigma V > 7$. Therefore, there is no reason to suggest that 
there is a zero mode contribution to the $\rho$ meson correlators being 
studied.


\section{Extrapolation results}
\label{sect:results}

\subsection{Renormalization flow curves}
\label{subsect:curves}

%

%

\begin{figure}[tp]
\includegraphics[height=0.80\hsize,angle=90]{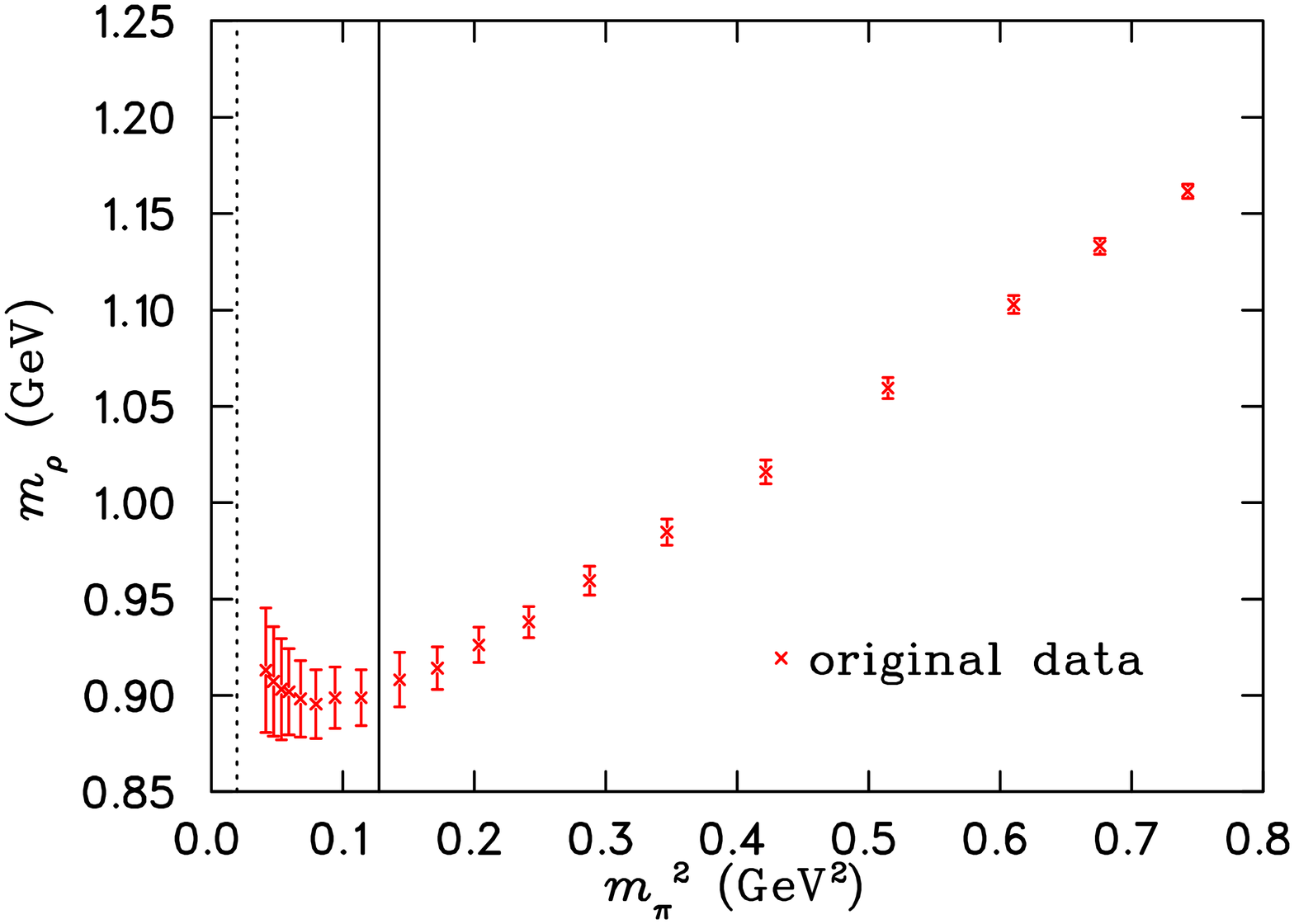}
\vspace{-11pt}
\caption{\footnotesize{(color online). Quenched lattice QCD data for
    the $\rho$ meson mass. The dashed
    vertical line indicates the physical pion mass and the solid
    vertical line shows how the data set is split into two parts. 
   The lower-mass portion of the data was not known at the time of 
   extrapolation. }}
\label{fig:data}
\vspace{5pt}
\includegraphics[height=0.80\hsize,angle=90]{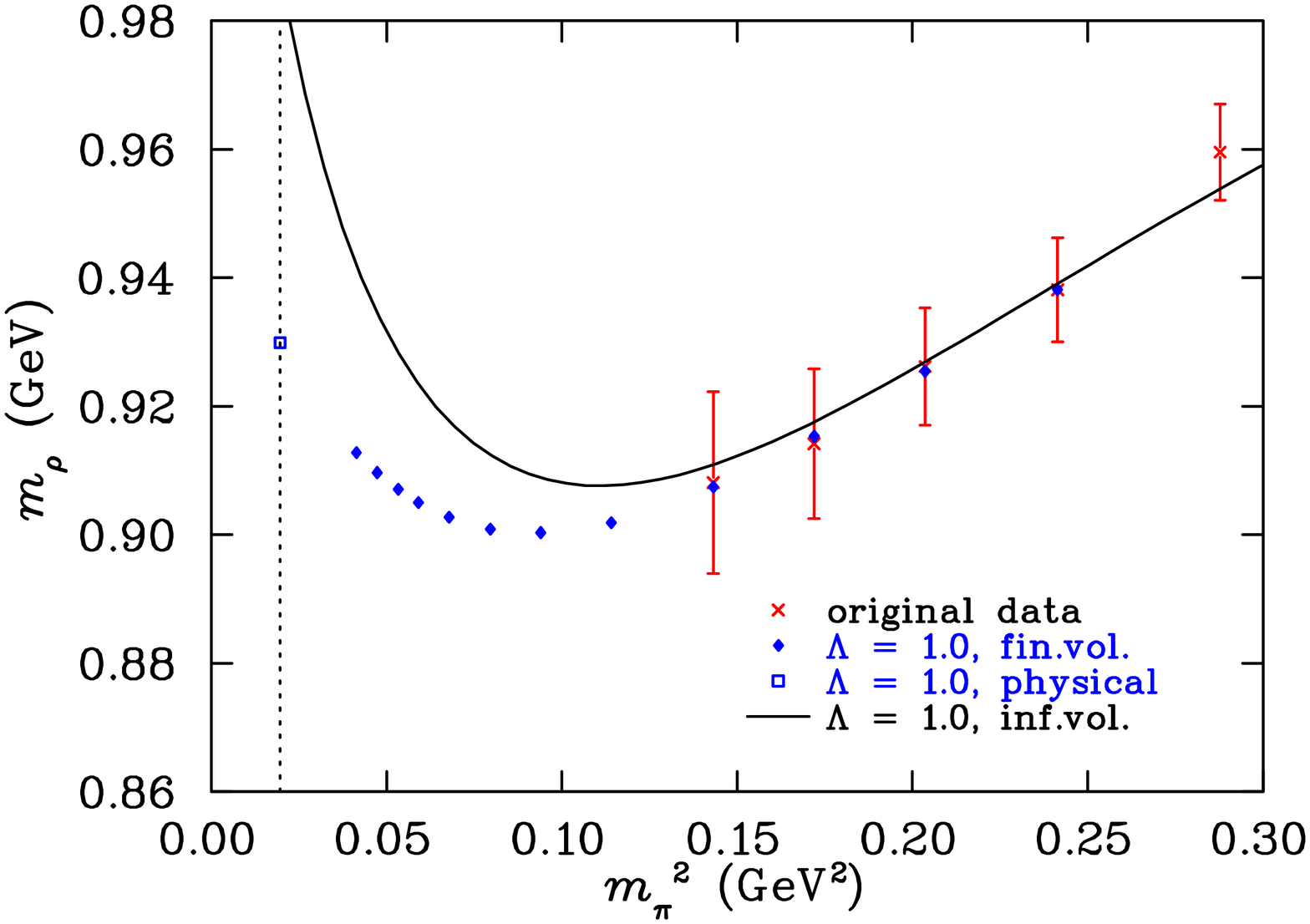}
\vspace{-11pt}
\caption{\footnotesize{(color online). A test extrapolation based on the four lightest original data points (excluding the low-energy set) as shown.  Both the finite- and infinite-volume results are shown for a triple-dipole regulator at $\La = 1.0$ GeV. The dashed vertical line indicates the physical pion mass.}}
\label{fig:testextrap}
\end{figure}

%
In order to produce an extrapolation to each test value of $m_\pi^2$,
 a finite-range regularization scale $\La$ must be selected.
As an example, one can choose a triple-dipole regulator at $\La = 1.0$ GeV.
 By using Eq.(\ref{eqn:renormexpsnfin}), finite- and infinite-volume 
extrapolations are shown in Figure 
\ref{fig:testextrap}. Note that the $m_\pi^2$ values selected for the 
 finite-volume extrapolations 
exactly correspond to the `missing' low-energy data points set aside
 earlier.  
The physical point $m_\pi^2 = 0.0196$ GeV$^2$ is included as well.
%

\begin{figure}[tp]
\includegraphics[height=0.80\hsize,angle=90]{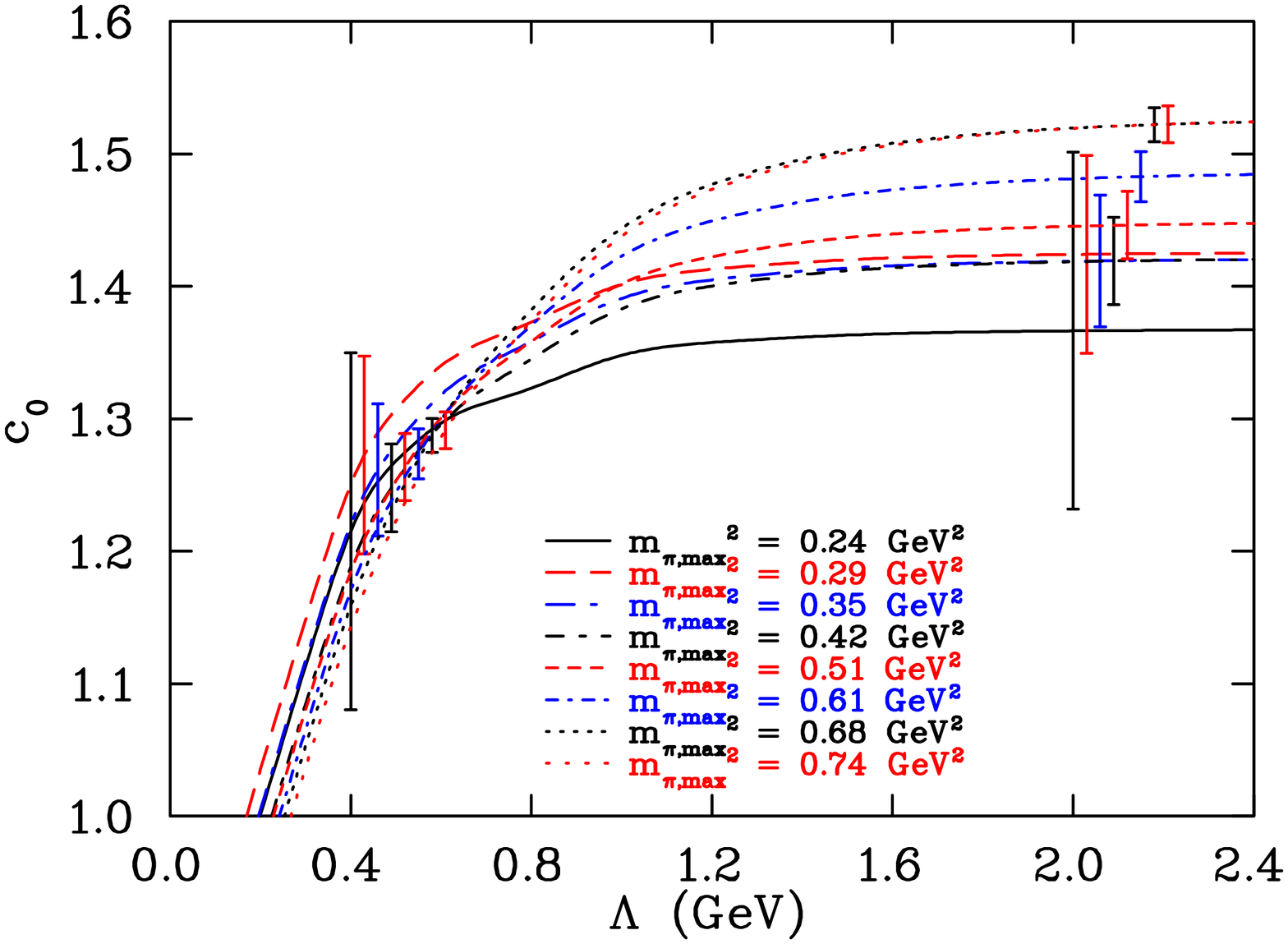}
\vspace{-11pt}
\caption{\footnotesize{(color online). Behaviour of $c_0$ vs.\ $\La$. A few points are selected to indicate the general size of the statistical error bars.}}
\label{fig:Kehfeic0}
\vspace{5pt}
\includegraphics[height=0.80\hsize,angle=90]{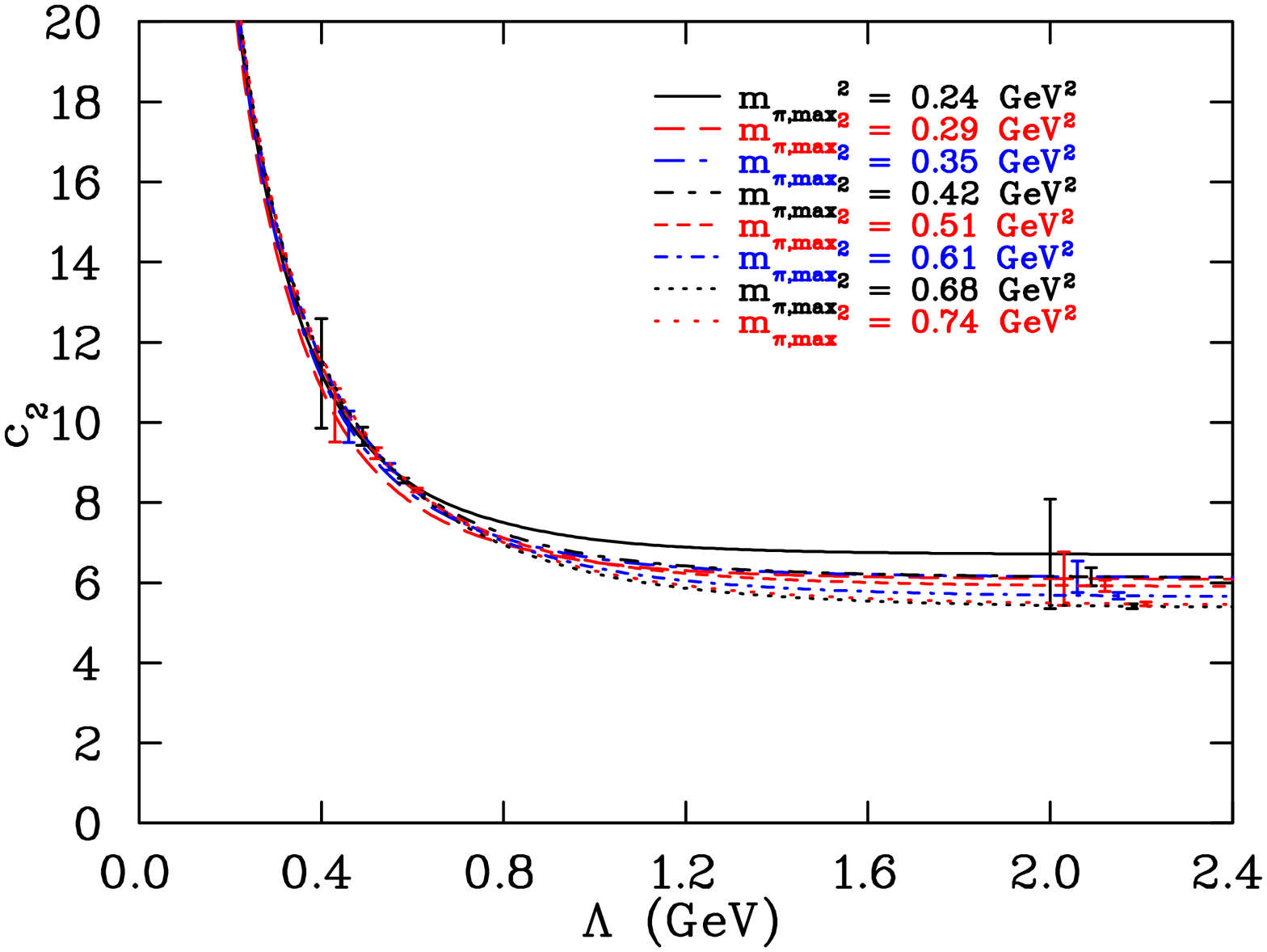}
\vspace{-11pt}
\caption{\footnotesize{(color online). Behaviour of $c_2$ vs.\ $\La$. A few points are selected to indicate the general size of the statistical error bars.}}
\label{fig:Kehfeic2}
\vspace{5pt}
\includegraphics[height=0.80\hsize,angle=90]{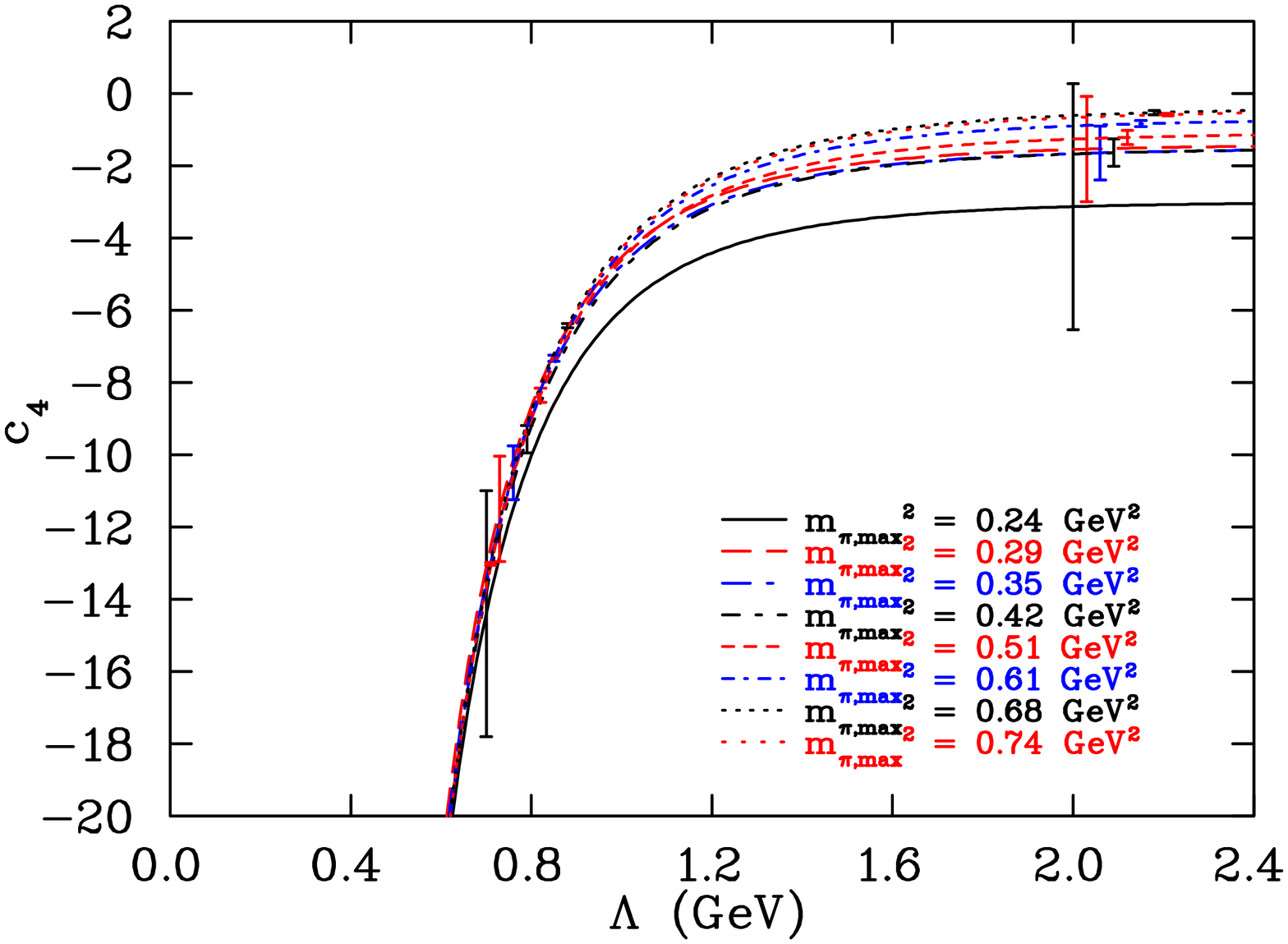}
\vspace{-11pt}
\caption{\footnotesize{(color online). Behaviour of $c_4$ vs.\ $\La$. A few points are selected to indicate the general size of the statistical error bars.}}
\label{fig:Kehfeic4}
\end{figure}

Now the regularization 
scale-dependence of low-energy coefficients $c_0$, $c_2$ and 
$c_4$ is investigated for various upper limits of the range of pion masses.
The renormalization of these low-energy coefficients  
 is considered for a series of $\La$ values.
The aim is to obtain renormalization flow curves, each  
corresponding to a
different value of maximum pion mass, $m_{\pi,\ro{max}}^2$. 
Thus the behaviour of the renormalization of the low-energy coefficients  
can be examined as lattice data extend further outside the PCR.
Figures \ref{fig:Kehfeic0} through \ref{fig:Kehfeic4}
 show the renormalization flow curves for each of $c_0$, $c_2$ and $c_4$.
Note that each data
point plotted has an associated error bar, but for the sake of clarity 
only a few points are selected to indicate the general size of the 
statistical error bars.
Using the procedure described in Ref.~\cite{Hall:2010ai},
 the optimal regularization scale 
 is identified by the value of $\La$ that minimizes the discrepancies among 
the renormalization flow curves. This 
indicates the value of regularization scale 
 at which the renormalization of $c_0$, $c_2$ 
and $c_4$ is least sensitive to the truncation of the data. Physically, 
this value of $\La$ can be associated with an intrinsic scale related 
to the size of the source of the pion cloud. 
%

%

By examining Figures \ref{fig:Kehfeic0} through \ref{fig:Kehfeic4}, 
 increasing $m_{\pi,\ro{max}}^2$ 
 leads to greater scheme-dependence in the 
renormalization, since the data sample lies 
further from the PCR. 
Complete scheme-independence would be indicated by a horizontal line at the 
 physical point.
%
 Since the effective field theory is calculated to a 
finite chiral order, complete scheme-independence across all possible 
  values of $\La$ will not occur in practice.
 Note that an asymptotic value is usually  
observed in the renormalization flow as $\La$ becomes large, indicating that 
 the higher-order terms of the chiral expansion are effectively zero. 
 However, these asymptotic values of the low-energy coefficients 
are  poor estimates 
of their correct values, as previously demonstrated in a pseudodata 
model \cite{Hall:2010ai}. Instead, the best estimates of the low-energy
 coefficients lie in the identification of the intersection point of the 
renormalization flow of the low-energy coefficients.
It is also of note that, 
for small values of $\La$, FRR schemes break down. 
The regularization scale must be at least large enough to include
the chiral physics being studied. 
%


\subsection{Optimal regularization scale}
\label{subsect:err}

The optimal regularization scale $\La^\ro{scale}$ can be obtained from the 
renormalization flow curves using a chi-square analysis described below.
 In addition, 
 the analysis will allow the extraction of a range for 
$\La^\ro{scale}$. Knowing how the data are correlated, the systematic 
uncertainties from the coupling constants and $\La^\ro{scale}$ will be combined
 to obtain an error bar for each extrapolation point. 
Of particular interest are the values of $m_{\rho,Q}$ at the values of 
$m_\pi^2$ explored in the lattice simulations but excluded in the chiral 
extrapolation.

\begin{figure}[tp]
\includegraphics[height=0.80\hsize,angle=90]{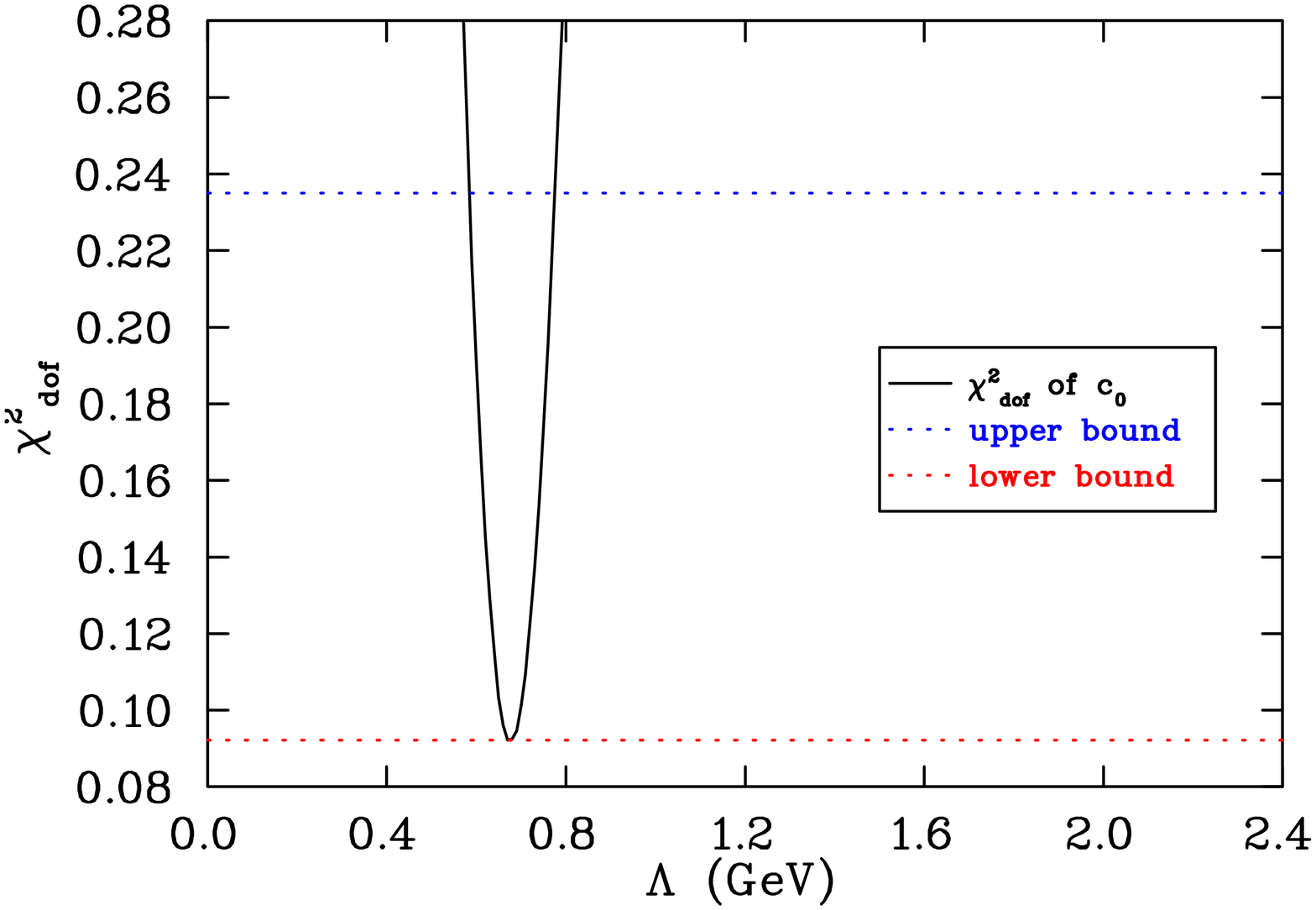}
\vspace{-11pt}
\caption{\footnotesize{(color online). $\chi^2_{dof}$ for $c_0$ versus $\La$, corresponding to the renormalization flow curves displayed in Figure \ref{fig:Kehfeic0}.}}
\label{fig:Kehfeichisqdofc0}
\vspace{5pt}
\includegraphics[height=0.80\hsize,angle=90]{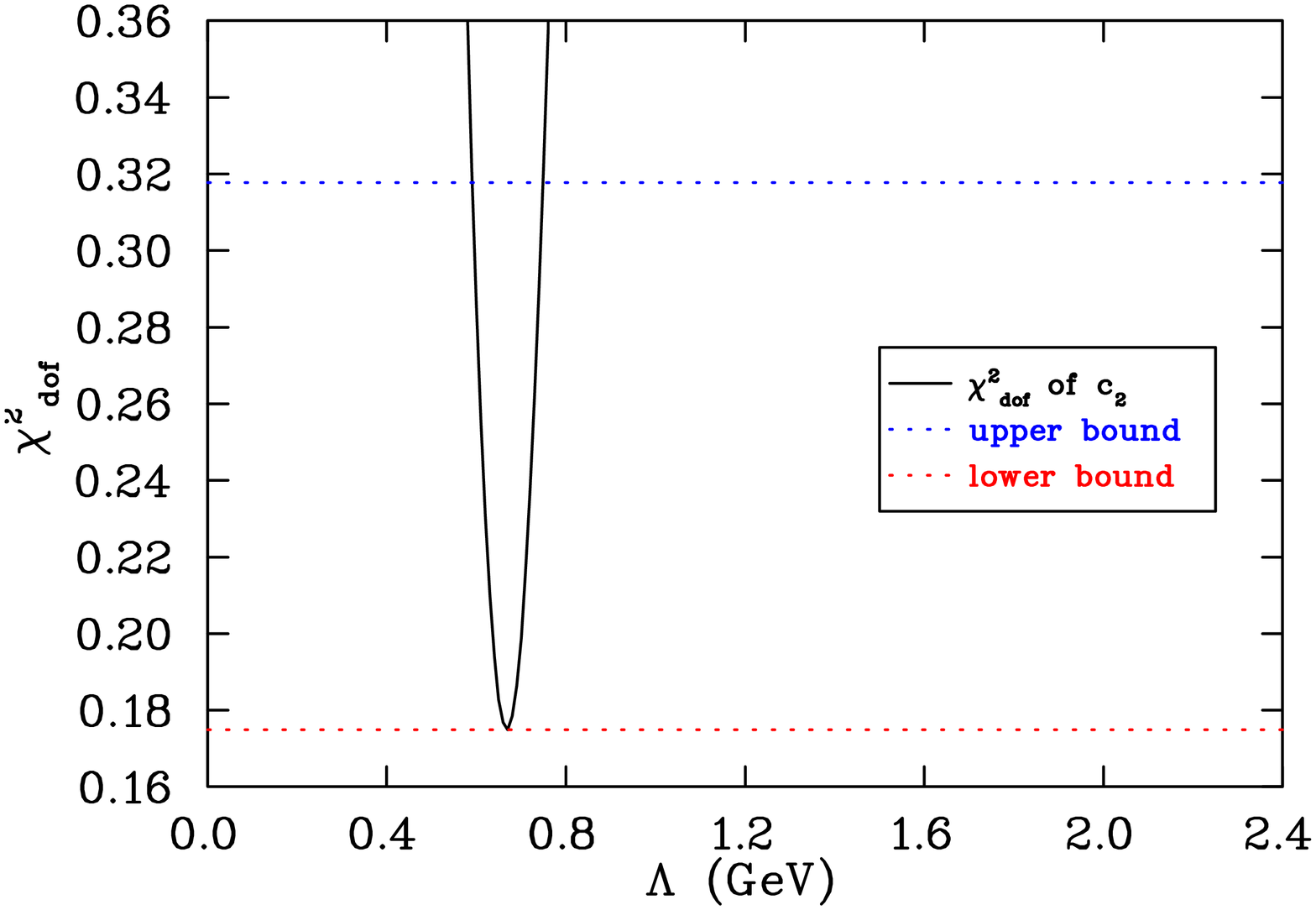}
\vspace{-11pt}
\caption{\footnotesize{(color online). $\chi^2_{dof}$ for $c_2$ versus $\La$, corresponding to the renormalization flow curves displayed in Figure \ref{fig:Kehfeic2}.}}
\label{fig:Kehfeichisqdofc2}
\vspace{5pt}
\includegraphics[height=0.80\hsize,angle=90]{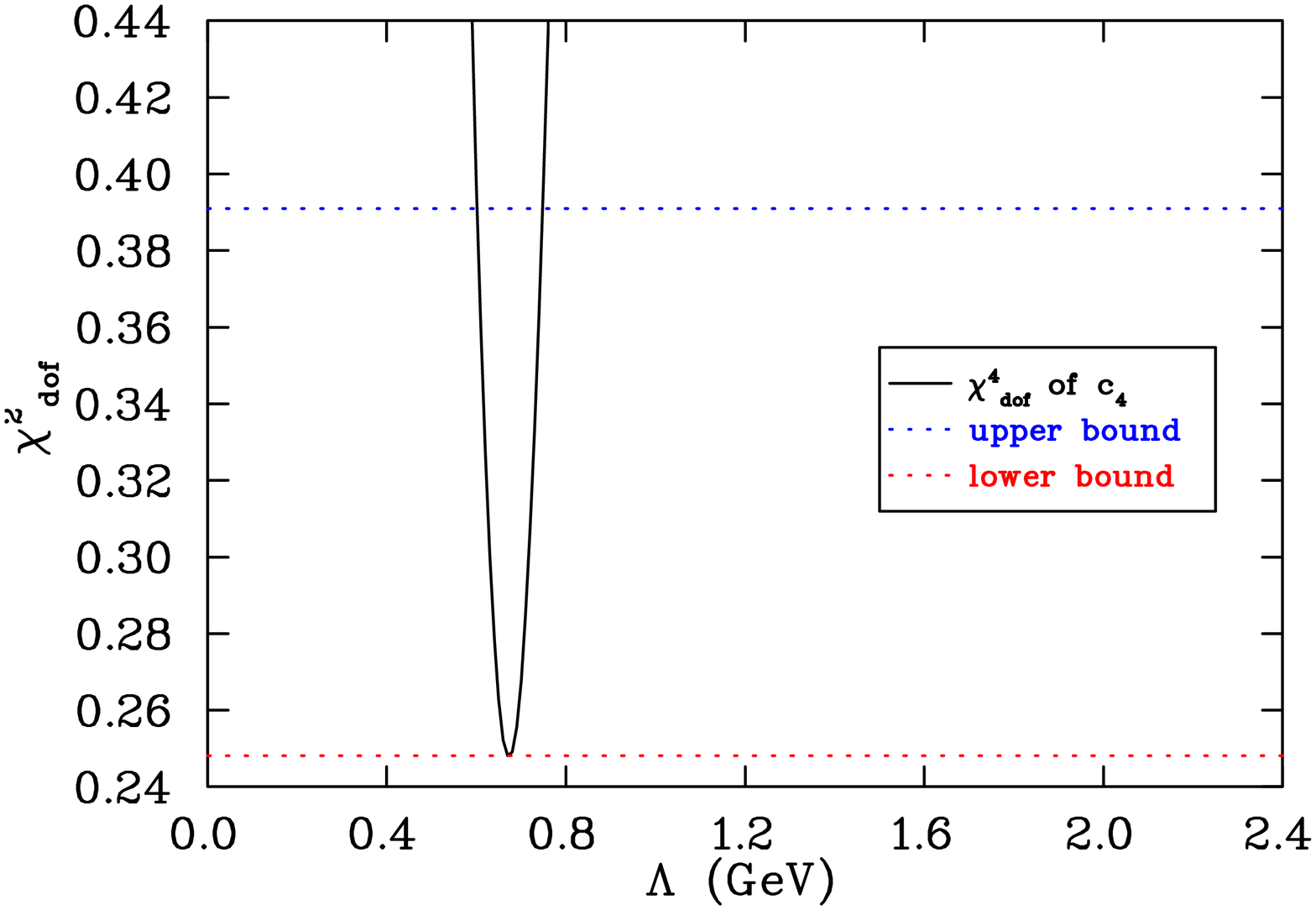}
\vspace{-11pt}
\caption{\footnotesize{(color online). $\chi^2_{dof}$ for $c_4$ versus $\La$, corresponding to the renormalization flow curves displayed in Figure \ref{fig:Kehfeic4}.}}
\label{fig:Kehfeichisqdofc4}
\end{figure}

To obtain a measure of the uncertainty associated with 
an optimal regularization scale, 
a $\chi^2_{dof}$ function is constructed. This function should allow 
 easy identification of the intersection points in the renormalization 
flow curves, and a range associated with this central regularization scale.
The first step is to plot $\chi^2_{dof}$
 against a series of $\La$ values. 
The relevant data 
are the 
extracted low-energy coefficients with differing values of 
$m_{\pi,\ro{max}}^2$.
 A plot of $\chi^2_{dof}$ is constructed separately
 for each renormalized coefficient $c$ (with uncertainty $\de c$):
\eqb
\chi^2_{dof} = \f{1}{n-1} \sum_{i=1}^{n} \f{{(c(i\,;\La) - c^T(\La))}^2}
{{(\de c(i\,;\La))}^2},
\eqe
for $i$ corresponding to 
 fits with differing values of 
$m_{\pi,\ro{max}}^2$ ($n = 8$). The theoretical value $c^T$ is given by the
weighted mean:
\eqb
c^T(\La) = \f{\sum_{i=1}^{n}c(i\,;\La)/{{(\de c(i\,;\La))}^2}}
{\sum_{j=1}^{n} 1 / {(\de c(j\,;\La))}^2}.
\eqe
The $\chi^2_{dof}$ plots using a triple-dipole regulator are shown in 
Figures \ref{fig:Kehfeichisqdofc0} through \ref{fig:Kehfeichisqdofc4}. 
The optimal regularization scale $\La^\ro{scale}$ 
is taken to be the central value $\La_\ro{central}$ of each plot. 
The upper and lower bounds obey 
 the condition $\chi^2_{dof} < \chi^2_{dof, min} + 1/(dof)$. 
The results for the optimal regularization scale and the 
 upper and lower bounds are shown in 
Table~\ref{table:scales}.
 It is remarkable that each low-energy coefficient leads to 
the same optimal value of $\La$, i.e. $\La_{\ro{central}} = 0.67$ GeV. 
 By averaging the results among $c_0$, $c_2$,
 and $c_4$, the optimal regularization scale $\La_{\ro{scale}}$
for the quenched $\rho$ meson mass can be calculated for this data set: 
 $\La_{\ro{scale}} = 0.67^{+0.09}_{-0.08}$ GeV. %

\begin{figure}[tp]
\begin{center}
\vspace{-5mm}
\includegraphics[height=0.80\hsize,angle=90]{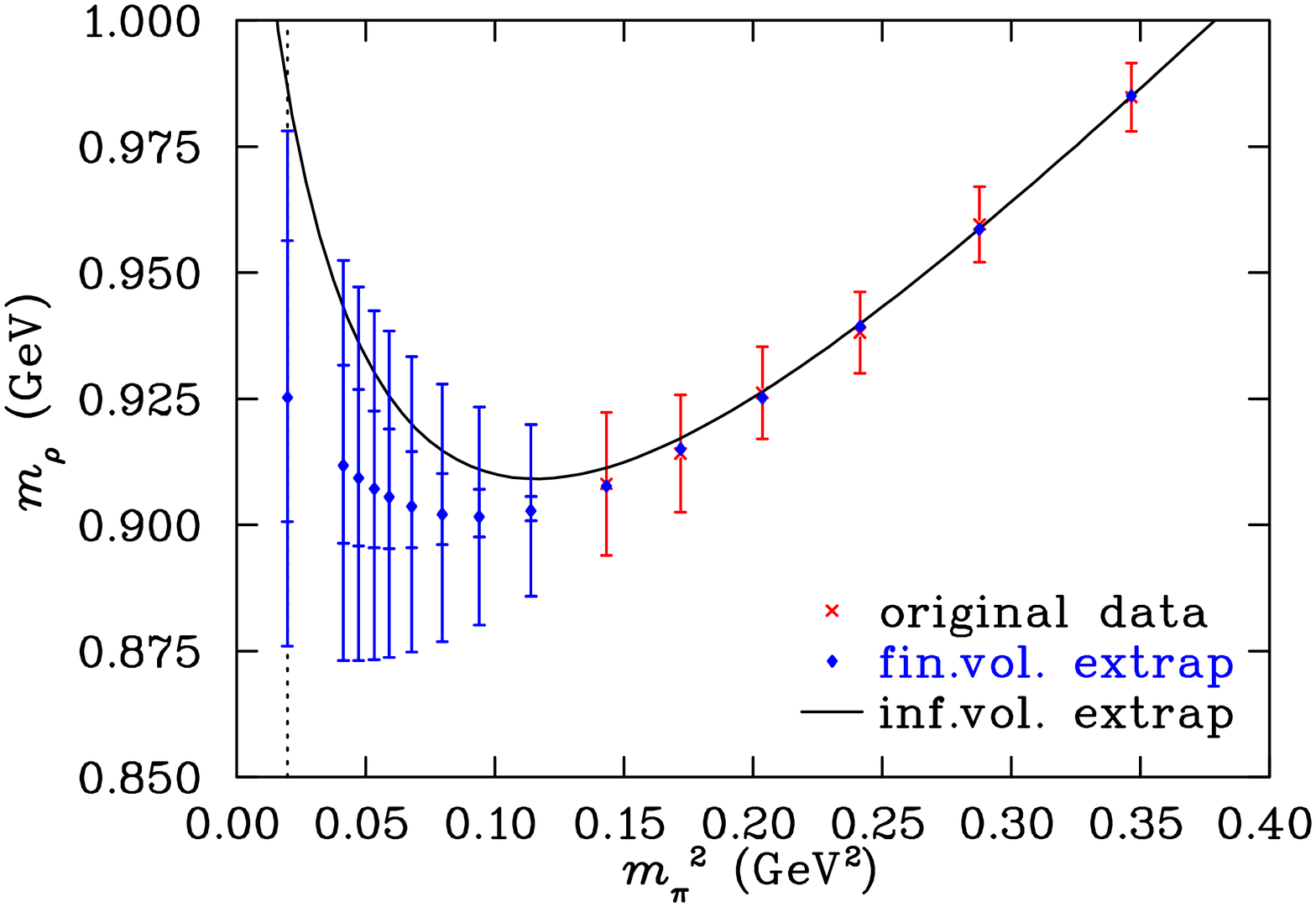}
\vspace{-11pt}
\caption{\footnotesize{(color online). Extrapolation at $\La_{\ro{scale}} = 0.67^{+0.09}_{-0.08}$ GeV based on Kentucky Group data, and using the optimal number of data points, corresponding to $\hat{m}_{\pi,\ro{max}}^2 = 0.35$ GeV$^2$. The inner error bar on the extrapolation points represents purely the systematic error from parameters. The outer error bar represents the systematic and statistical error estimates added in quadrature.}}
\label{fig:initialextrap}
%
\vspace{5pt}
\includegraphics[height=0.80\hsize,angle=90]{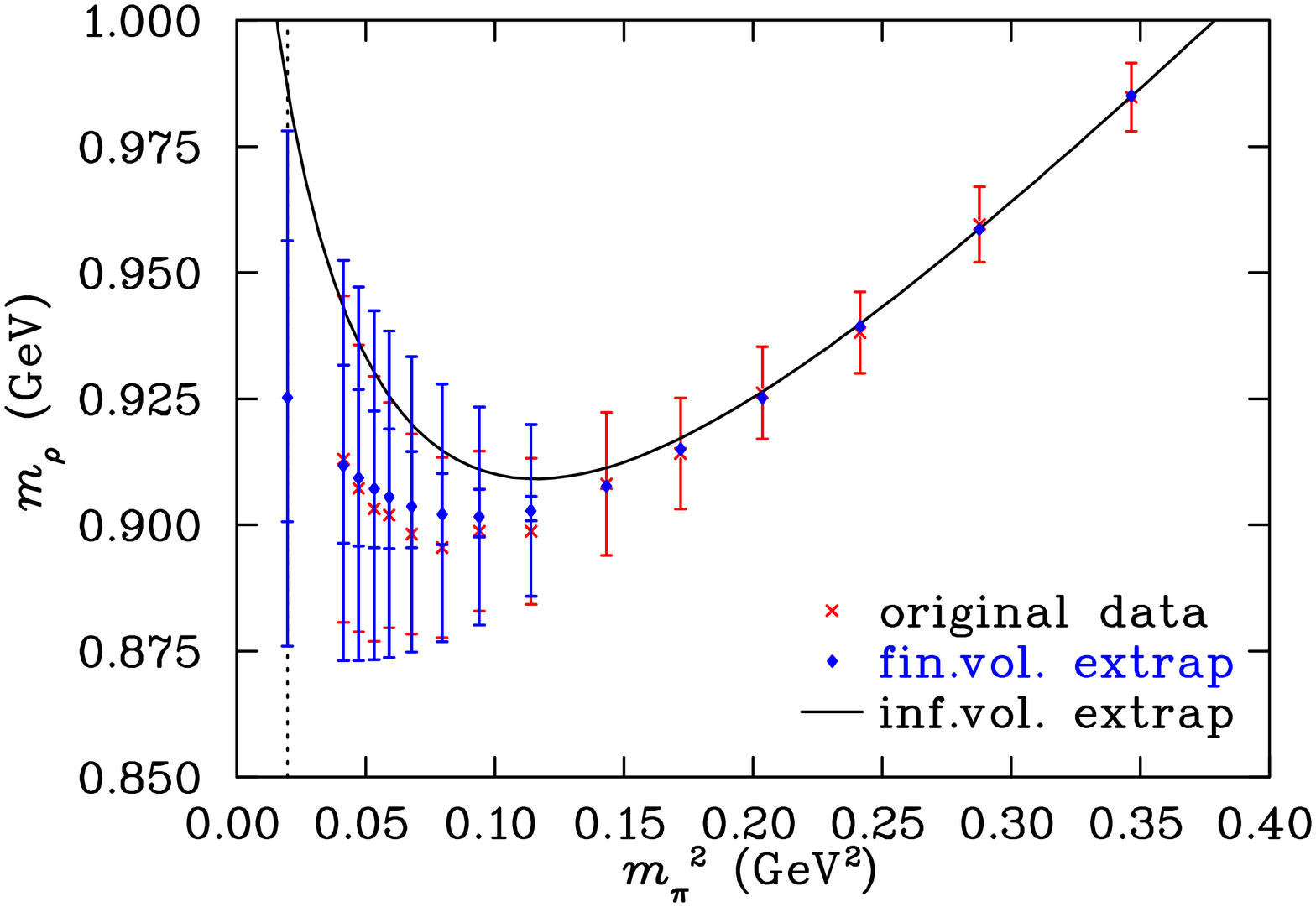}
\vspace{-11pt}
\caption{\footnotesize{(color online). Comparison of chiral extrapolation predictions (blue diamond) with Kentucky Group data (red cross).  Extrapolation is performed at $\La_{\ro{scale}} = 0.67^{+0.09}_{-0.08}$ GeV, and using the optimal number of data points, corresponding to $\hat{m}_{\pi,\ro{max}}^2 = 0.35$ GeV$^2$. The inner error bar on the extrapolation points represents purely the systematic error from parameters. The outer error bar represents the systematic and statistical error estimates added in quadrature.}}
\label{fig:finalextrap}
\includegraphics[height=0.80\hsize,angle=90]{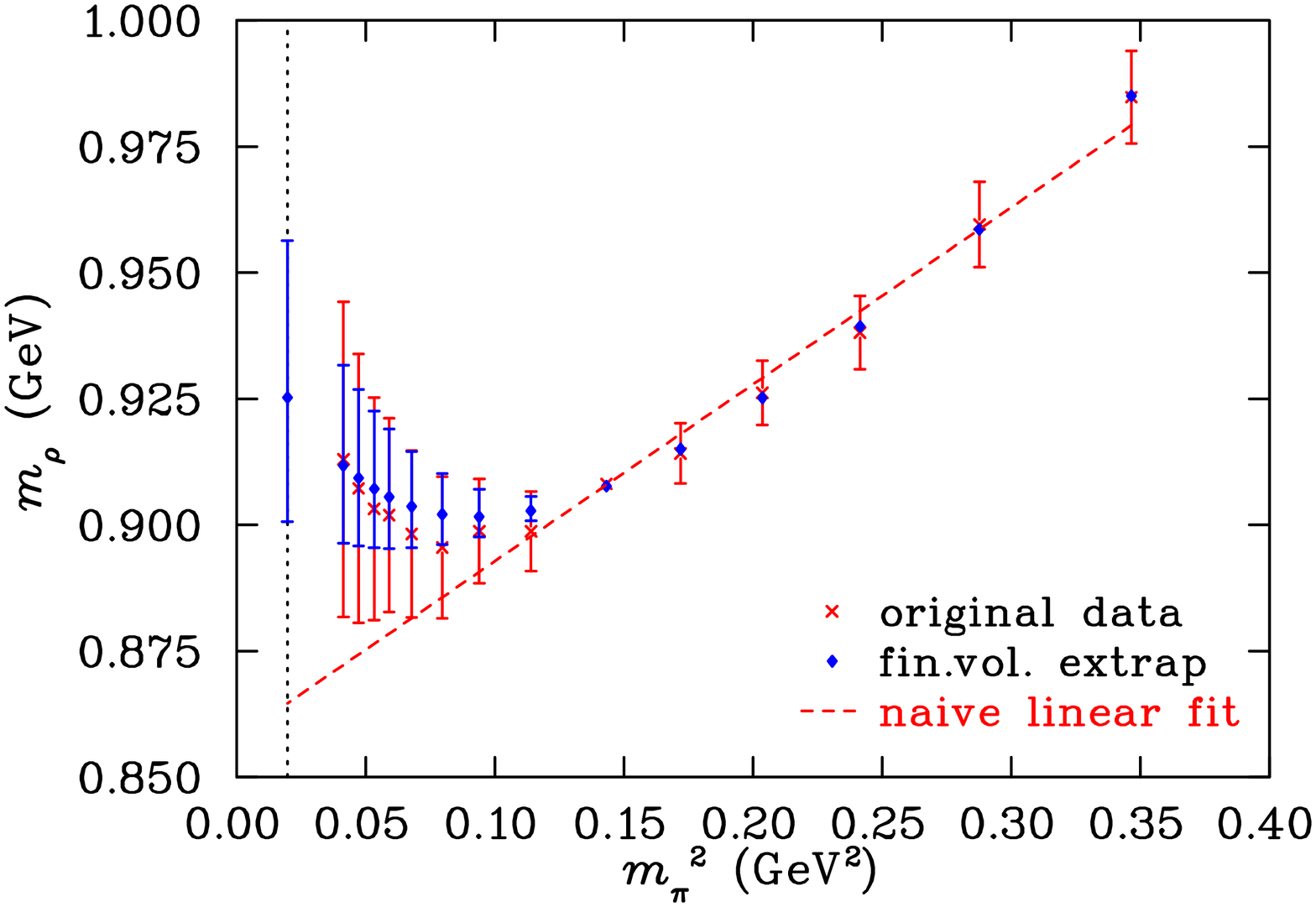}
\vspace{-11pt}
\caption{\footnotesize{(color online). Comparison of chiral extrapolation predictions (blue diamond) with Kentucky Group data (red cross), with errors correlated relative to the point at $m_\pi^2 = 0.143$ GeV$^2$.  This is done simply to clarify the plot in Figure \ref{fig:finalextrap} by removing much of the correlated statistical error. Extrapolation is performed at $\La_{\ro{scale}} = 0.67^{+0.09}_{-0.08}$ GeV, and using the optimal number of data points, corresponding to $\hat{m}_{\pi,\ro{max}}^2 = 0.35$ GeV$^2$. The error bar on the extrapolation points represents the systematic error only. A simple linear fit, on the optimal pion mass region, is included for comparison.}}
\label{fig:finaldeltaextrap}
\end{center}
\end{figure}

The result of the final extrapolation, using the estimate of the optimal 
regularization scale 
$\La_{\ro{scale}} = 0.67^{+0.09}_{-0.08}$ GeV, and using the initial data set
to predict the low-energy data points, is shown in Figure 
\ref{fig:initialextrap}. 
The extrapolation to the physical point obtained for this quenched data set is: 
$m_{\rho,Q}^{\ro{ext}}(m_{\pi,\ro{phys}}^2) = 0.925^{+0.053}_{-0.049}$ GeV,  
an uncertainty 
of less than $6$\%.

\begin{table}[tp]
 \caption{\footnotesize{Values of the central, upper and lower regularization scales, in GeV, obtained from the $\chi^2_{dof}$ analysis of $c_0$, $c_2$ and $c_4$, displayed in Figures \ref{fig:Kehfeichisqdofc0} through \ref{fig:Kehfeichisqdofc4}.}}
  \newcommand\T{\rule{0pt}{2.8ex}}
  \newcommand\B{\rule[-1.4ex]{0pt}{0pt}}
  \begin{center}
    \begin{tabular}{llll}
      \hline
      \hline
       \T\B 
      scale (GeV) \qquad & $c_0$ (Fig.\ref{fig:Kehfeichisqdofc0}) $\,\,\,\,\,$& $c_2$ (Fig.\ref{fig:Kehfeichisqdofc2}) $\,\,$& $c_4$ (Fig.\ref{fig:Kehfeichisqdofc4})   \\
      \hline
      $\La_\ro{central}$  &\T $0.67$ & $0.67$ & $0.67$ \\
      $\La_\ro{upper}$   &\T $0.78$ & $0.75$ & $0.75$ \\
      $\La_\ro{lower}$   &\T $0.58$ & $0.59$ & $0.60$ \\
      \hline
    \end{tabular}
  \end{center}
\vspace{-6pt}
  \label{table:scales}
\end{table}

Note that each extrapolation point  
displays two error bars.
 The inner error bar corresponds to the 
systematic uncertainty 
in the parameters only, 
 and the outer error bar corresponds
to the systematic and statistical uncertainties 
of each point added in quadrature.
Also, the infinite-volume extrapolation curve is displayed in order to 
illustrate the effect of finite-volume corrections to the loop integrals.

In Figure \ref{fig:finalextrap}, 
the extrapolation predictions are compared 
against the actual simulation results, which were not included in the fit. 
Note 
that both the extrapolations and the simulation results display the 
same non-analytic curvature near the physical point. 
%
Figure \ref{fig:finaldeltaextrap} shows the data plotted with error bars 
correlated relative to the lightest data point in the original set, 
$m_\pi^2 = 0.143$ GeV$^2$. 
This aids in clarifying the plot from Figure \ref{fig:finalextrap}
 by removing much of the correlated 
statistical error in the lattice data, and allows us to be even 
more stringent in determining whether the extrapolation is successful. 
 It is notable that the extrapolated results are consistent with 
the lattice data even after having removed the correlated statistical error. 
To highlight the importance of this application of an extended 
$\chi$EFT, 
a simple linear fit is included in Figure \ref{fig:finaldeltaextrap}.  
 By ignoring low-energy chiral physics, the 
linear fit is statistically incorrect at the physical point.
 Note also that 
all of the missing original data points are consistent within 
the extrapolations' 
systematic uncertainties. 
%
After 
 statistical correlations are subtracted, the extrapolated points correspond to
 an 
error bar almost half 
the size of that of the lattice data points. 
 In order to match this precision at low energies, the time
 required in lattice simulations would increase by approximately four times.

\begin{figure}[tp]
\includegraphics[height=0.80\hsize,angle=90]{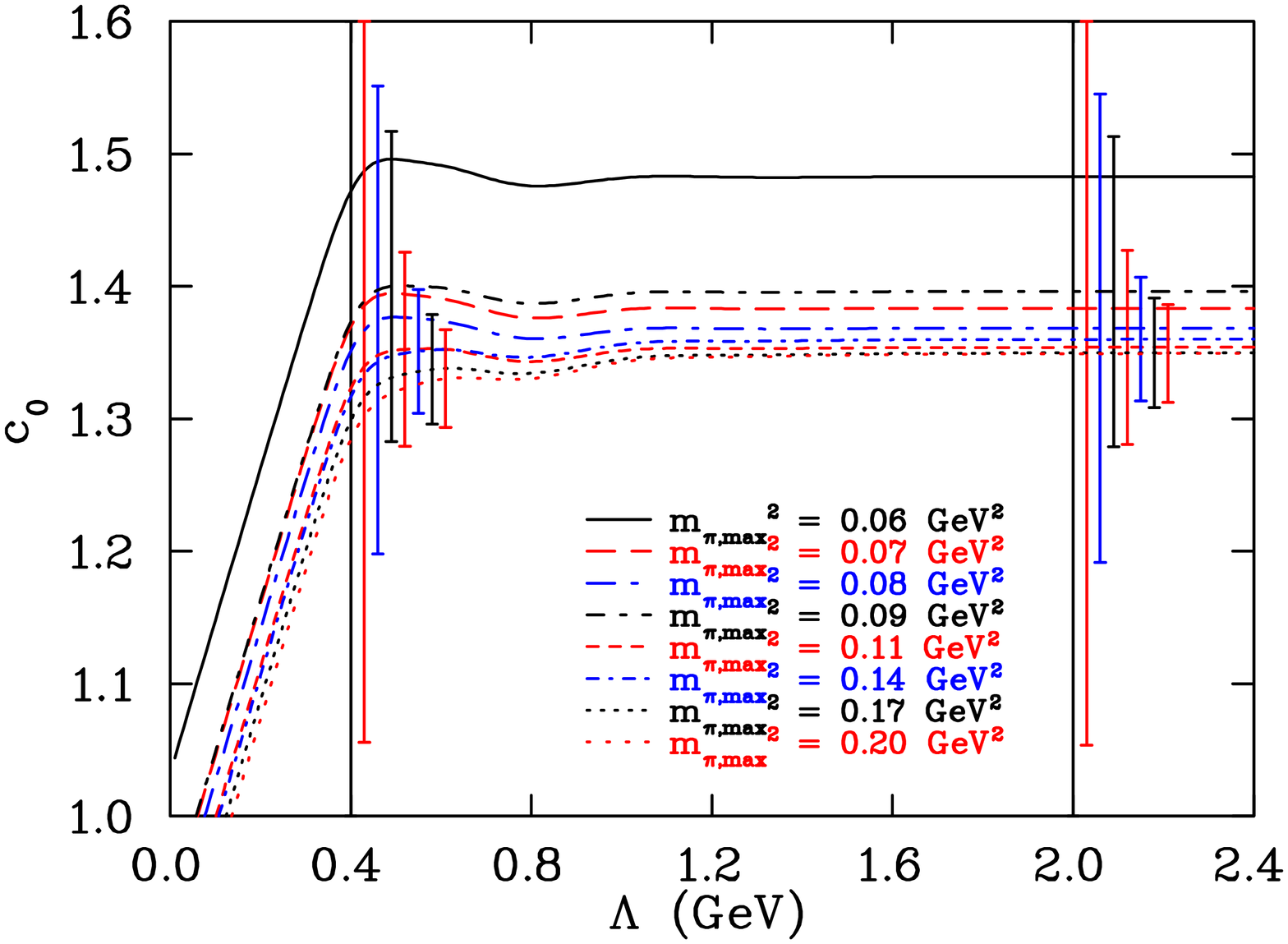}
\vspace{-11pt}
\caption{\footnotesize{(color online). Behaviour of $c_0$ vs.\ $\La$ including the initially excluded low-energy data. A few points are selected to indicate the general size of the statistical error bars.}}
\label{fig:Kehfeic0new}
\vspace{5pt}
\includegraphics[height=0.80\hsize,angle=90]{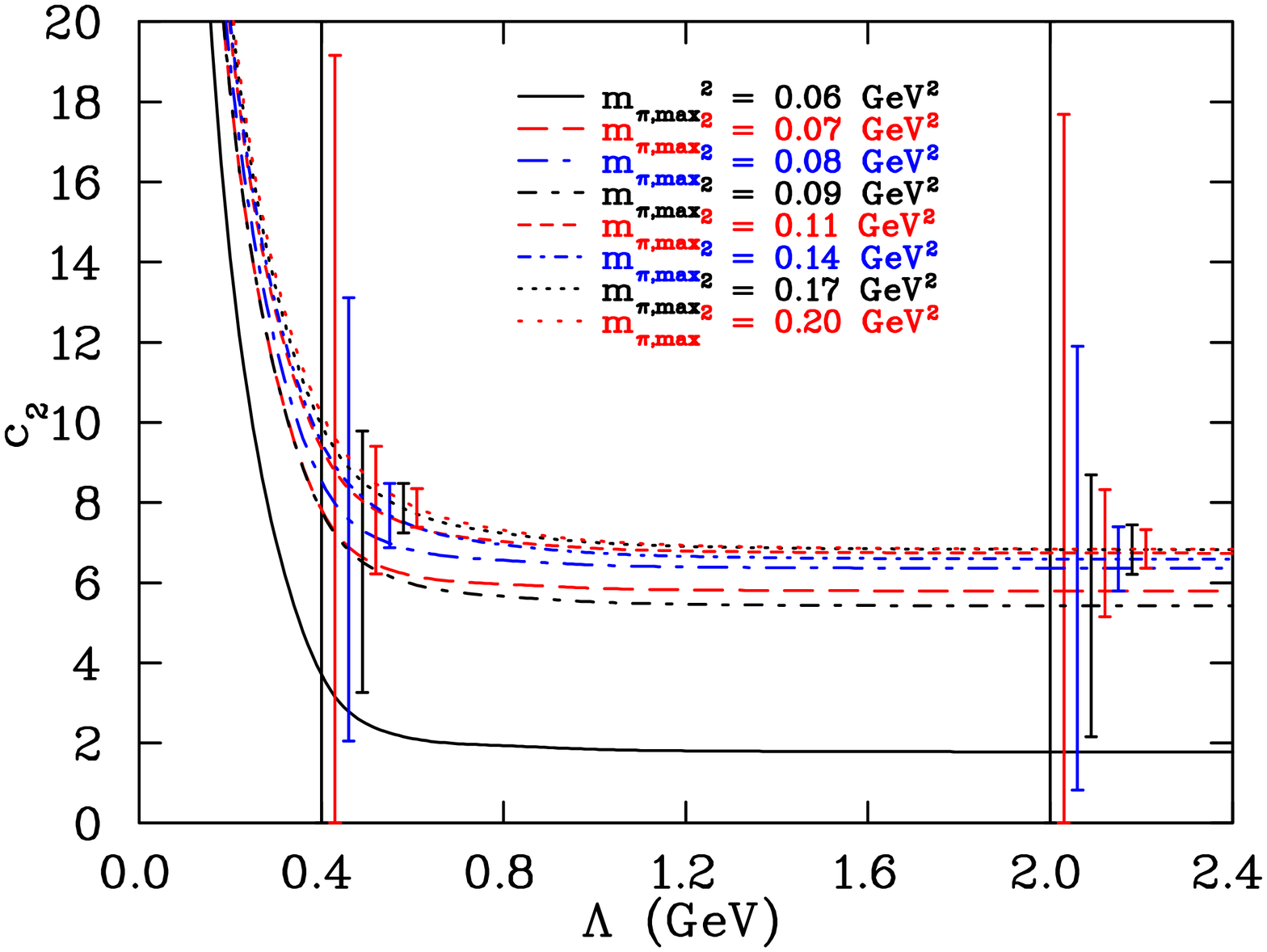}
\vspace{-11pt}
\caption{\footnotesize{(color online). Behaviour of $c_2$ vs.\ $\La$ including the initially excluded low-energy data. A few points are selected to indicate the general size of the statistical error bars.}}
\label{fig:Kehfeic2new}
\vspace{5pt}
\includegraphics[height=0.80\hsize,angle=90]{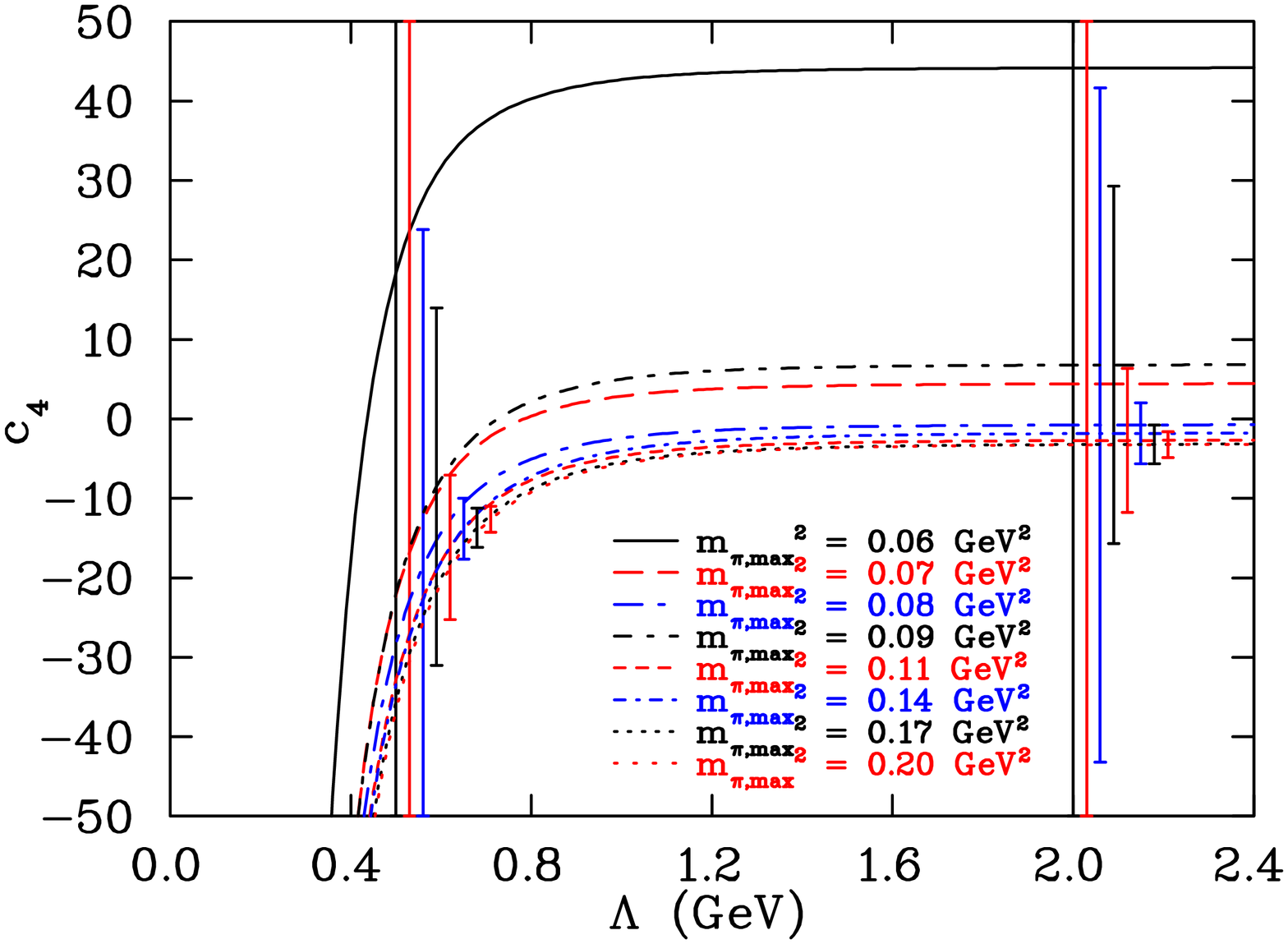}
\vspace{-11pt}
\caption{\footnotesize{(color online). Behaviour of $c_4$ vs.\ $\La$ including the initially excluded low-energy data. A few points are selected to indicate the general size of the statistical error bars.}}
\label{fig:Kehfeic4new}
\end{figure}

\begin{figure}[tp]
\includegraphics[height=0.80\hsize,angle=90]{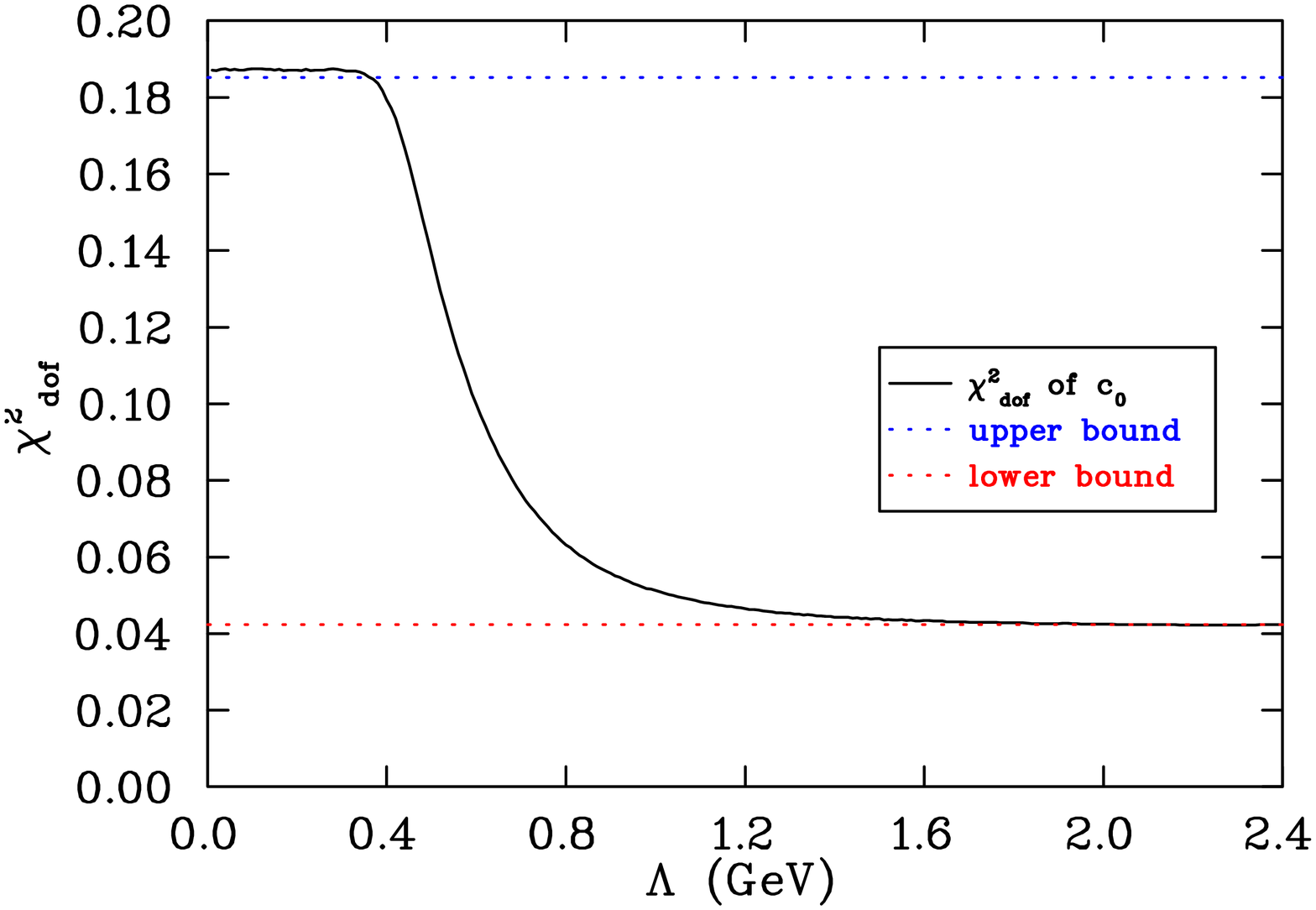}
\vspace{-11pt}
\caption{\footnotesize{(color online). $\chi^2_{dof}$, for $c_0$ versus $\La$, corresponding to the renormalization flow curves displayed in Figure \ref{fig:Kehfeic0new}. A lower bound for the regularization scale is found: $\La^{c_0}_{\ro{lower}} = 0.39$ GeV.}}
\label{fig:Kehfeichisqdofc0new}
\vspace{5pt}
\includegraphics[height=0.80\hsize,angle=90]{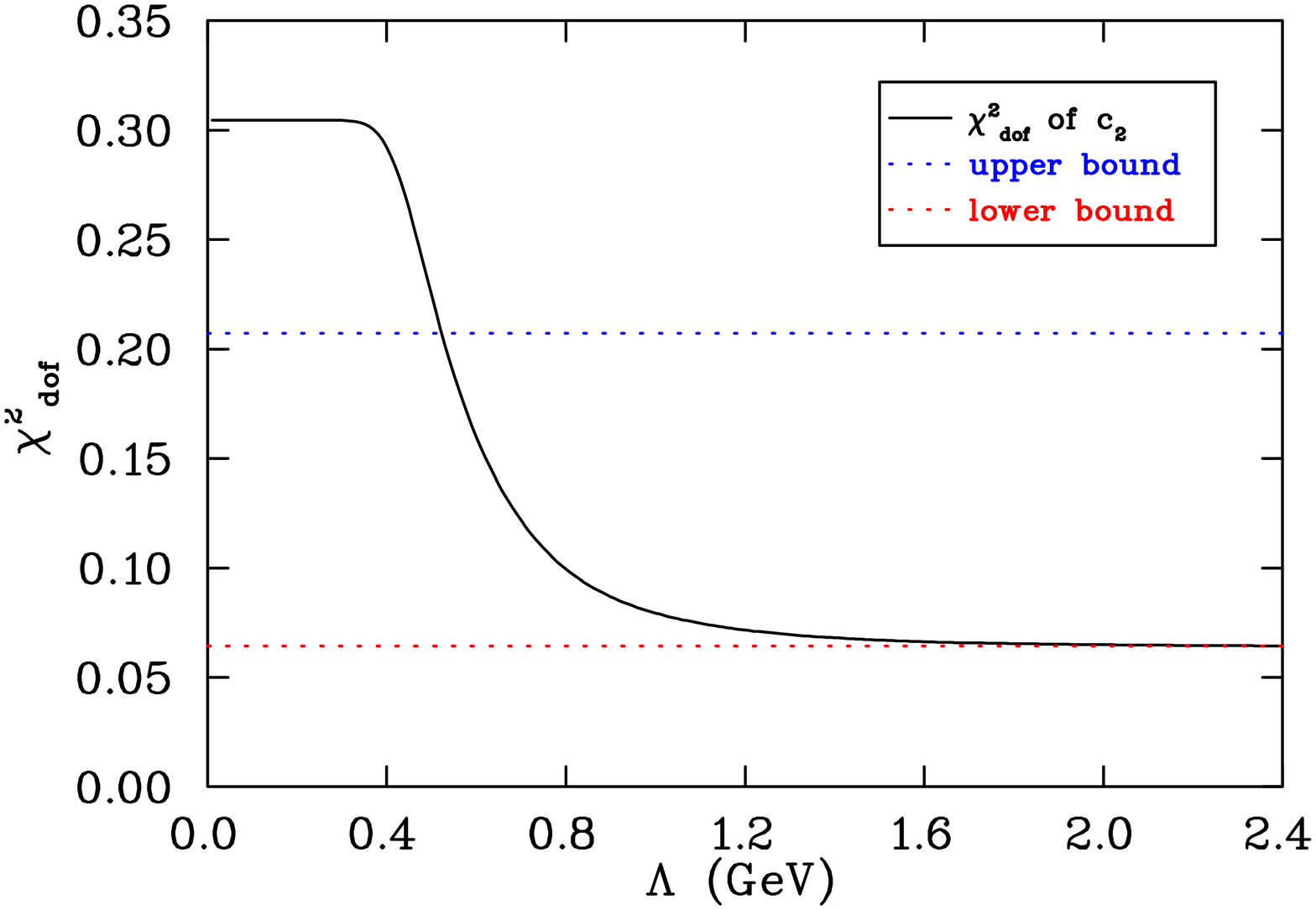}
\vspace{-11pt}
\caption{\footnotesize{(color online). $\chi^2_{dof}$, for $c_2$ versus $\La$, corresponding to the renormalization flow curves displayed in Figure \ref{fig:Kehfeic2new}. A lower bound for the regularization scale is found: $\La^{c_2}_{\ro{lower}} = 0.52$ GeV.}}
\label{fig:Kehfeichisqdofc2new}
\vspace{5pt}
\includegraphics[height=0.80\hsize,angle=90]{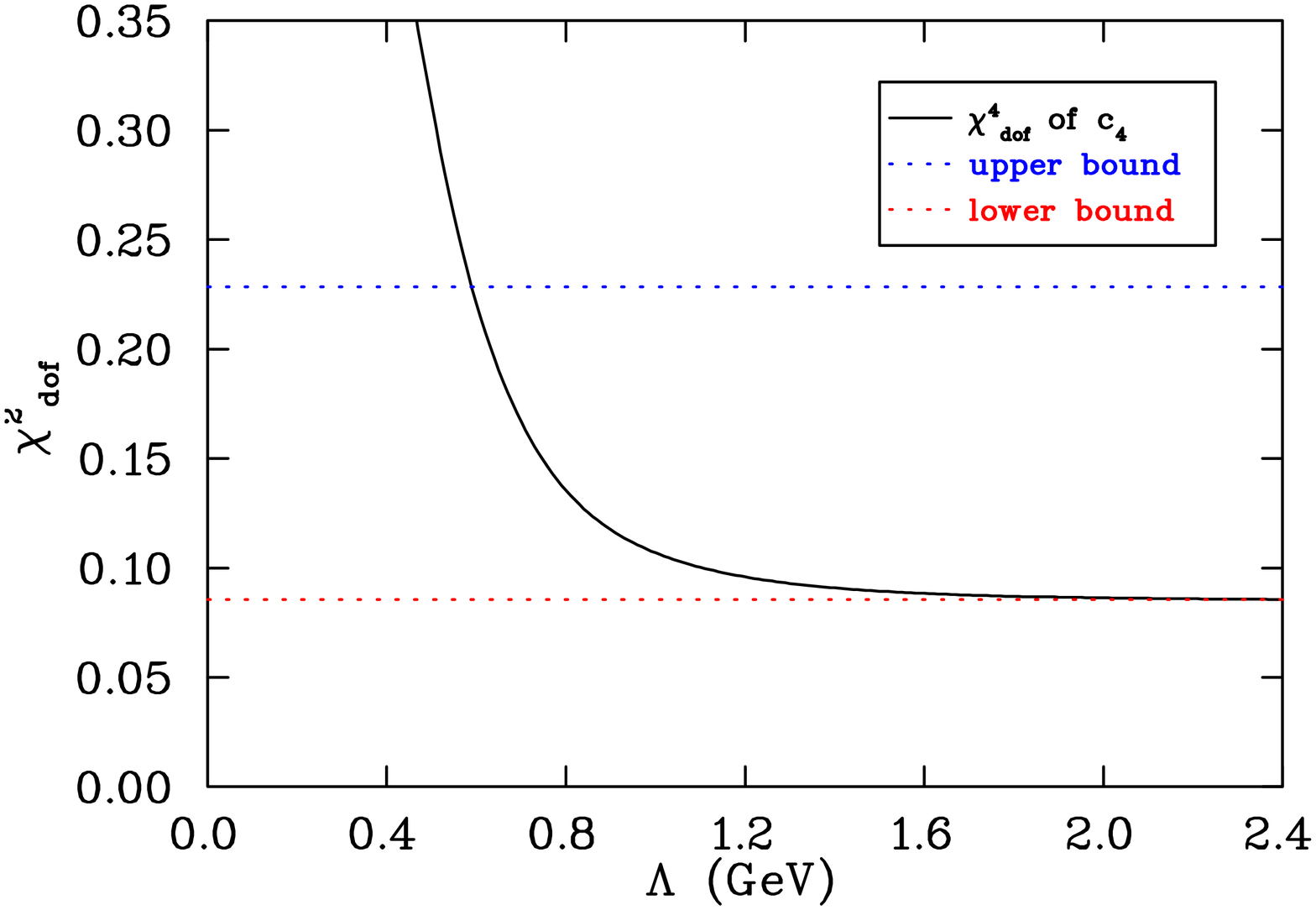}
\vspace{-11pt}
\caption{\footnotesize{(color online). $\chi^2_{dof}$, for $c_4$ versus $\La$, corresponding to the renormalization flow curves displayed in Figure \ref{fig:Kehfeic4new}. A lower bound for the regularization scale is found: $\La^{c_4}_{\ro{lower}} = 0.59$ GeV.}}
\label{fig:Kehfeichisqdofc4new}
\end{figure}

In order to check if scheme-independence is recovered using data within the 
PCR, the low-energy data that were initially excluded from analysis 
can now be treated in the same way. That is, renormalization flow curves can 
be constructed as a function of $\La$ 
for sequentially increasing $m_{\pi,\ro{max}}^2$. The results are 
shown in Figures \ref{fig:Kehfeic0new} through \ref{fig:Kehfeic4new}. Clearly, 
the renormalization flow curves for each plot corresponding to $c_0$, $c_2$ 
and $c_4$ are flatter than those of the initial analysis, 
indicating a reduction in the regularization 
scale-dependence due to the use of data closer to the PCR. 
One is not able to extract an optimal regularization scale from these 
plots, as shown in the behaviour of $\chi^2_{dof}$, displayed  
in Figures \ref{fig:Kehfeichisqdofc0new} through \ref{fig:Kehfeichisqdofc4new}. 
However, each $\chi^2_{dof}$ curve provides a lower bound for the regularization 
scale, where FRR breaks down \cite{Hall:2010ai}, 
as discussed in Section \ref{subsect:curves}. 
These lower bounds are: $\La^{c_0}_{\ro{lower}} = 0.39$ GeV, 
$\La^{c_2}_{\ro{lower}} = 0.52$ GeV and $\La^{c_4}_{\ro{lower}} = 0.59$ GeV.  

\begin{figure}[tp]
\includegraphics[height=0.80\hsize,angle=90]{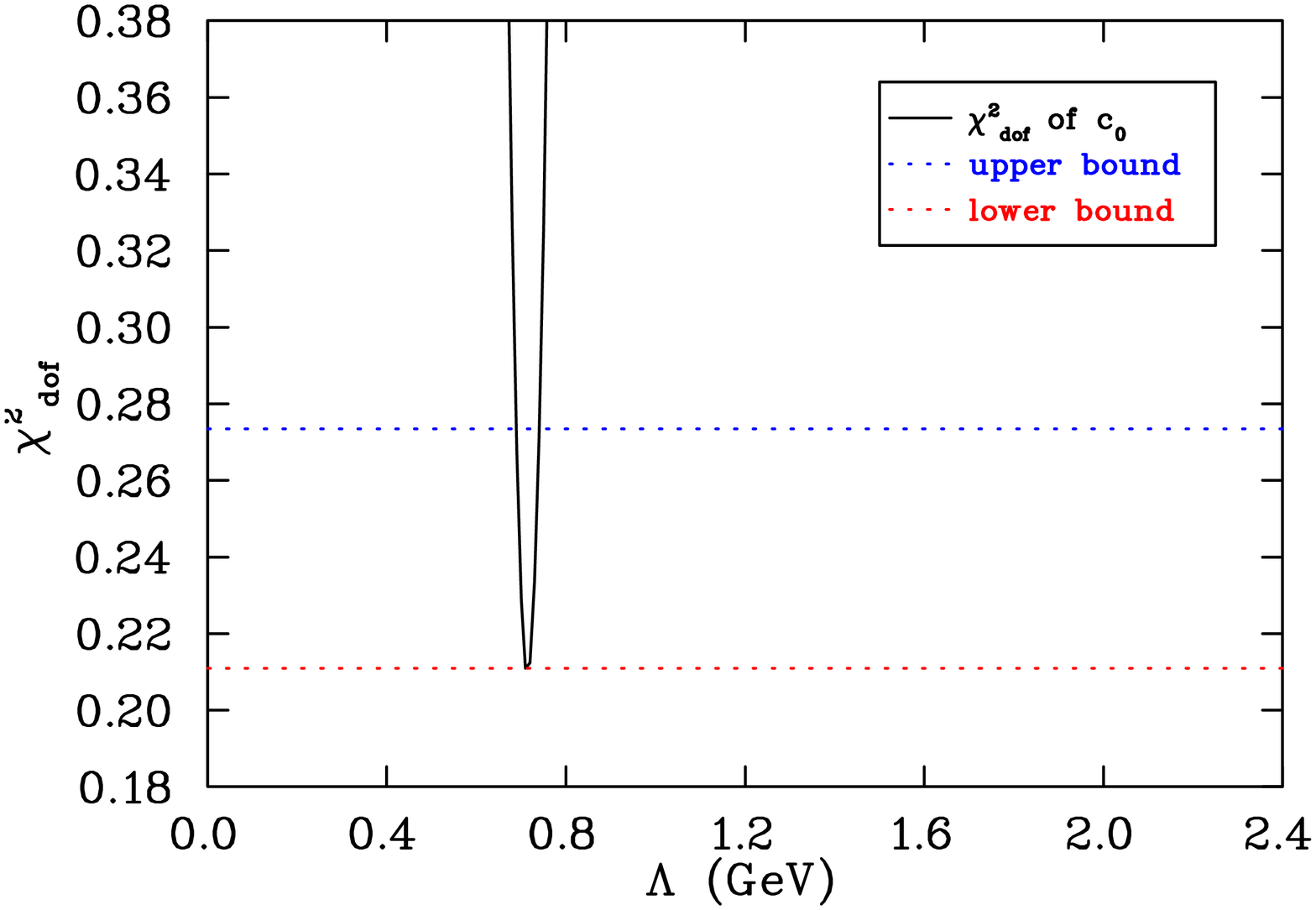}
\vspace{-11pt}
\caption{\footnotesize{(color online). $\chi^2_{dof}$, for $c_0$ versus $\La$, corresponding to all available data, including the low-energy set.}}
\label{fig:Kehfeichisqdofc0new17}
\vspace{5pt}
\includegraphics[height=0.80\hsize,angle=90]{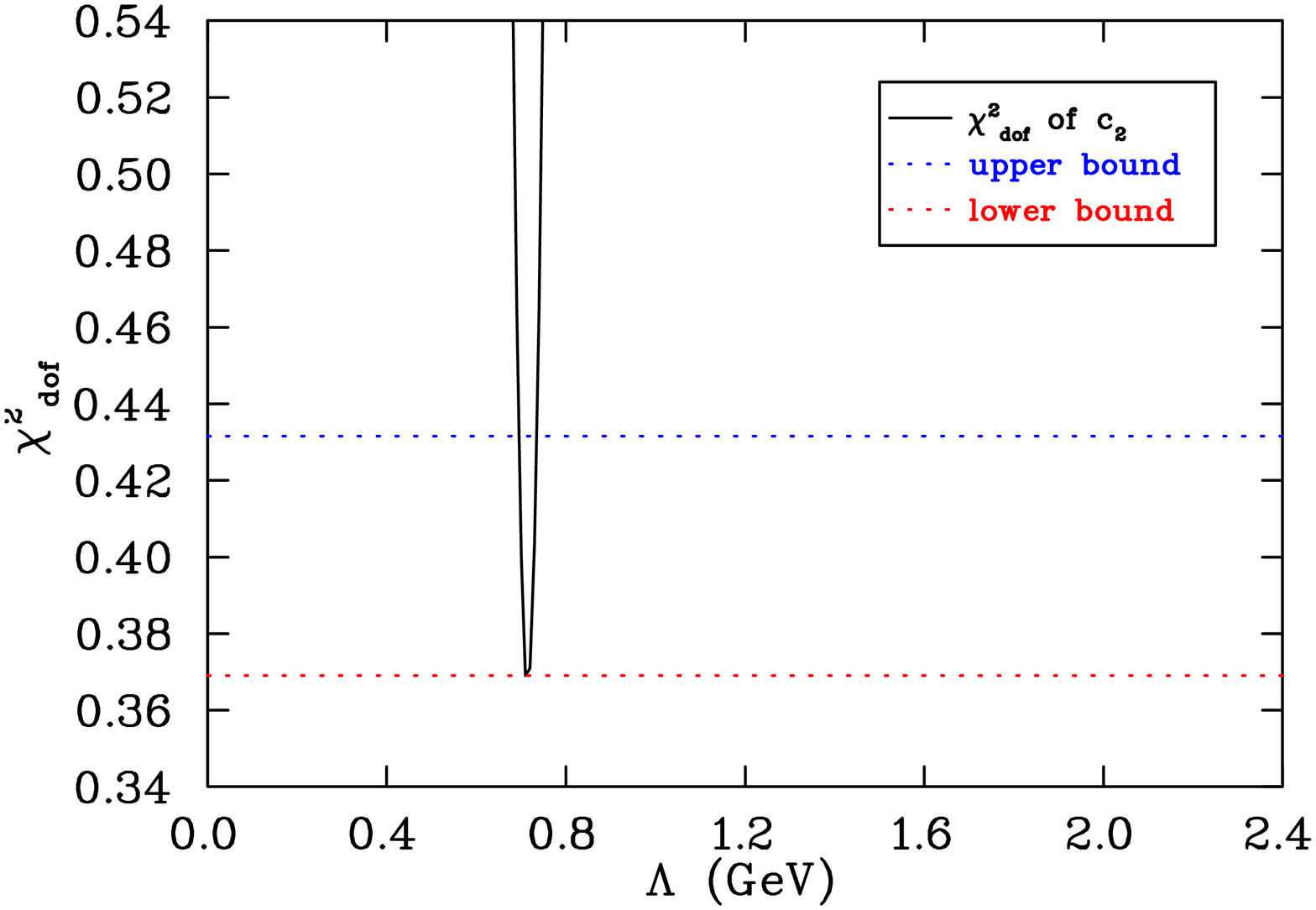}
\vspace{-11pt}
\caption{\footnotesize{(color online). $\chi^2_{dof}$, for $c_2$ versus $\La$, corresponding to all available data, including the low-energy set.}}
\label{fig:Kehfeichisqdofc2new17}
\vspace{5pt}
\includegraphics[height=0.80\hsize,angle=90]{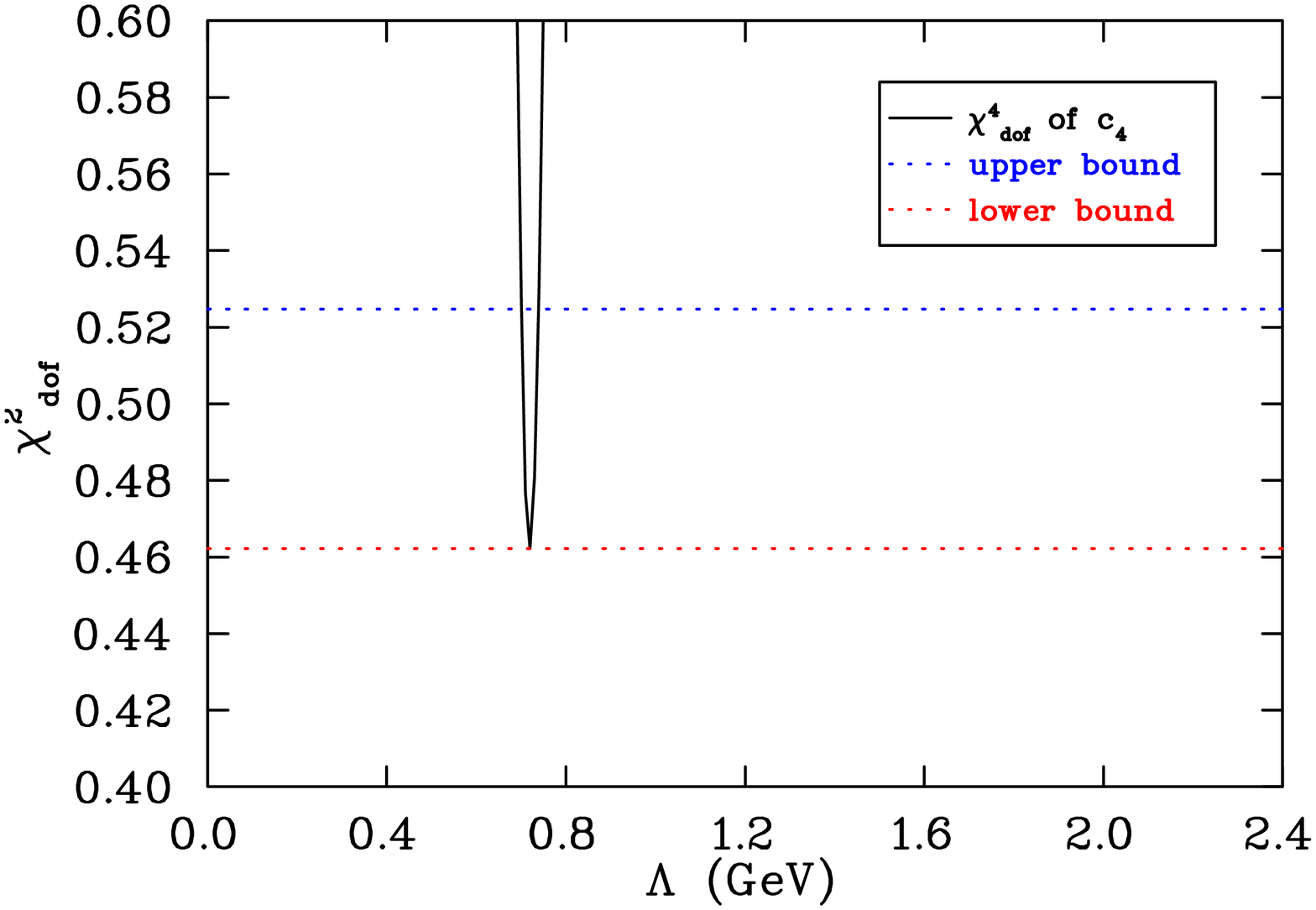}
\vspace{-11pt}
\caption{\footnotesize{(color online). $\chi^2_{dof}$, for $c_4$ versus $\La$, corresponding to all available data, including the low-energy set.}}
\label{fig:Kehfeichisqdofc4new17}
\end{figure}

The statistical error bars of the low-energy coefficients 
 corresponding to a small number of data points in 
Figures \ref{fig:Kehfeic0new} through \ref{fig:Kehfeic4new}  is large, 
and a statistical 
difference among the curves does not appear until 
$m_{\pi,\ro{max}}^2 \approx 0.11$ GeV$^2$. 
Thus the identification of an optimal regularization scale will be aided by  
incorporating data corresponding to even larger values of $m_{\pi,\ro{max}}^2$. 
By considering \emph{all} of the available data, the behaviour of 
$\chi^2_{dof}$, 
 as displayed in Figures \ref{fig:Kehfeichisqdofc0new17} through 
\ref{fig:Kehfeichisqdofc4new17}, resolve precise optimal regularization 
scales: $\La^{c_0}_\ro{central} = 0.72$ GeV, $\La^{c_2}_\ro{central} = 0.71$ GeV 
and $\La^{c_4}_\ro{central} = 0.71$ GeV. 
The systematic errors obtained 
from each $\chi^2_{dof}$ curve seem arbitrarily constrained as a consequence 
of including more data points, 
which extend well outside the chiral regime, and possibly 
outside the applicable region of FRR techniques. This issue is addressed 
in the ensuing section.

\subsection{Optimal pion mass region}
\label{subsect:err2}

\begin{figure}[tp]
\includegraphics[height=0.80\hsize,angle=90]{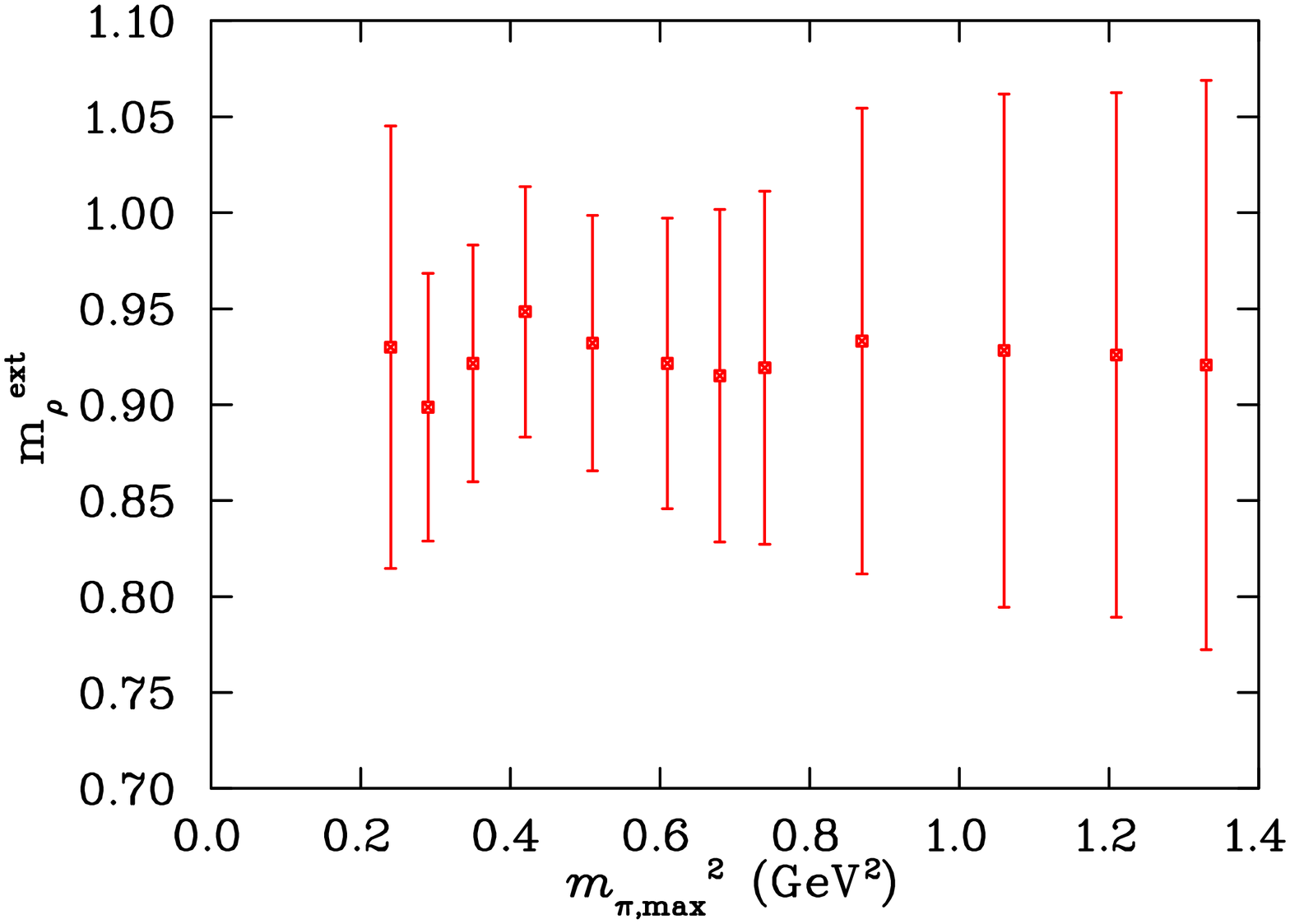}
\vspace{-11pt}
\caption{\footnotesize{(color online). Behaviour of the extrapolation of the quenched $\rho$ meson mass to the physical point $m_{\rho,Q}^{\ro{ext}}(m_{\pi,\ro{phys}}^2)$ vs.\ $m_{\pi,\ro{max}}^2$ using the initial data set, which excludes the lowest mass data points. In each case, $c_0$ is obtained using the scale $\La_{\ro{central}}$ (for a triple-dipole regulator) as obtained from the $\chi^2_{dof}$ analysis. The error bars include the statistical and systematic uncertainties in $c_0$ added in quadrature. The optimal value $\hat{m}_{\pi,\ro{max}}^2 = 0.35$ GeV$^2$.}}
\label{fig:Kehfeimrhovsmpisq}
\vspace{5pt}
\includegraphics[height=0.80\hsize,angle=90]{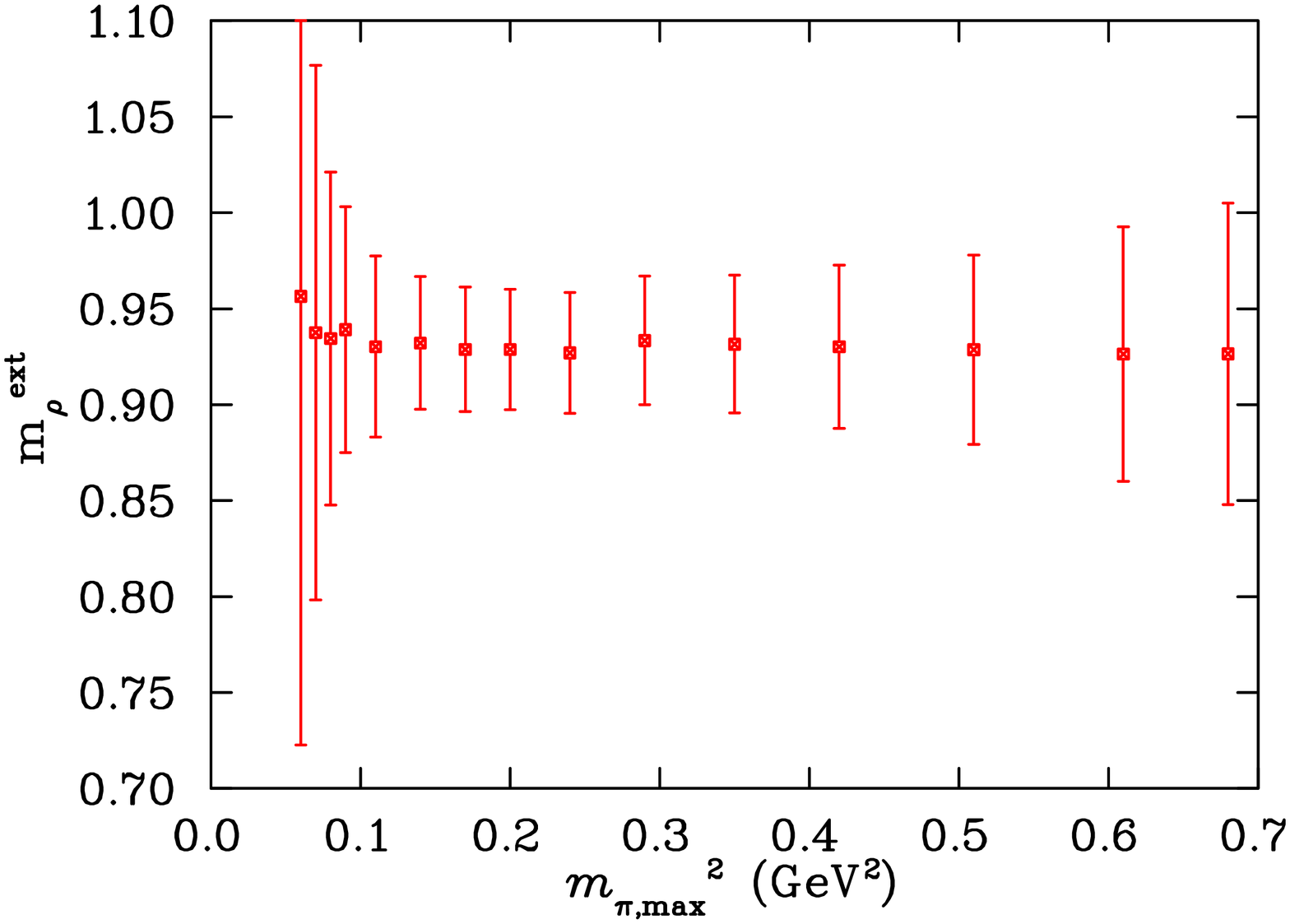}
\vspace{-11pt}
\caption{\footnotesize{(color online). Behaviour of the extrapolation of the quenched $\rho$ meson mass to the physical point $m_{\rho,Q}^{\ro{ext}}(m_{\pi,\ro{phys}}^2)$ vs.\ $m_{\pi,\ro{max}}^2$ using the complete data set, which includes the lowest mass data points. In each case, $c_0$ is obtained using the scale $\La_{\ro{central}}$ (for a triple-dipole regulator) as obtained from the $\chi^2_{dof}$ analysis. The error bars include the statistical and systematic uncertainties in $c_0$ added in quadrature. The optimal value $\hat{m}_{\pi,\ro{max}}^2 = 0.20$ GeV$^2$.}}
\label{fig:Kehfeimrhovsmpisqnew}
\end{figure}

In this section,  a robust method for determining an optimal range of 
pion masses is presented. This range
corresponds to an optimal number of simulation results to be used for fitting. 
First, consider 
the extrapolation
of the quenched $\rho$ meson mass, which can now be completed.
The statistical uncertainties in the values of $c_0$, $c_2$, $c_4$ 
are dependent on $m_{\pi,\ro{max}}^2$. As a consequence, the uncertainty in 
the extrapolated $\rho$ meson mass $m_\rho^{\ro{ext}}$ must also be 
dependent on $m_{\pi,\ro{max}}^2$. 
Since the estimate of the statistical uncertainty in an extrapolated point 
will tend to decrease as more data are included in the fit, one might 
na\"{i}vely choose to use the largest $m_{\pi,\ro{max}}^2$ value possible 
in the data set. However, at some large value of $m_{\pi,\ro{max}}^2$, 
FRR $\chi$EFT will not provide a valid model for obtaining a suitable fit. 
At this upper bound of applicability for FRR $\chi$EFT, the uncertainty 
in an extrapolated point is dominated by the systematic error in the 
underlying parameters. 
This is due to a greater scheme-dependence in extrapolations 
using data extending 
 outside the PCR, meaning that the extrapolations
    are more sensitive to changes in the parameters of the loop integrals. 
Thus there is a balance point $m_{\pi,\ro{max}}^2 \!= \hat{m}_{\pi,\ro{max}}^2$,  
where the statistical and systematic uncertainties (added in quadrature)
 in an extrapolation are minimized. 

In order to obtain this value $\hat{m}_{\pi,\ro{max}}^2$, consider 
the behaviour of the extrapolation of the $\rho$
 meson mass to the physical point $m_{\rho,Q}^{\ro{ext}}(m_{\pi,\ro{phys}}^2)$, 
as a function of $m_{\pi,\ro{max}}^2$. 
Treating the parameters: $\La^\ro{scale}$, $g_2$, $g_4$, $M_0^2$ and $A_0$ as 
independent, their systematic uncertainties from these sources are 
 added in quadrature. In addition, the systematic uncertainty due 
to the choice of the regulator functional form is roughly estimated 
by comparing the results using the double-dipole and the step function.  
These functional forms are the two most different forms of the 
various regulators considered, since the dipole was excluded due to 
the extra non-analytic contributions it introduces.
The results for the initial and 
complete data sets are shown in 
Figures \ref{fig:Kehfeimrhovsmpisq} and \ref{fig:Kehfeimrhovsmpisqnew}, 
respectively. Note that the systematic uncertainty due to $\La^\ro{scale}$ 
is included for chiral order $\ca{O}(m_\pi^4)$. 

  Figure \ref{fig:Kehfeimrhovsmpisq}
indicates an optimal value 
$\hat{m}_{\pi,\ro{max}}^2 = 0.35$ GeV$^2$, which will be used in the final 
extrapolations, in order to check the results of this method with the  
low-energy data. 
By using only the data contained in the optimal pion mass region, 
constrained by $\hat{m}_{\pi,\ro{max}}^2$, an estimate of 
the optimal regularization scale may be calculated with a 
more generous corresponding systematic 
uncertainty. 
The value $\La^\ro{scale} = 0.64$ GeV
is the average of $\La^\ro{scale}_{c_0}$, $\La^\ro{scale}_{c_2}$ and 
$\La^\ro{scale}_{c_4}$ using this method.  The $\chi^2_{dof}$ analysis 
does not provide an upper or lower bound at this value of 
$\hat{m}_{\pi,\ro{max}}^2$. Note that 
these two estimates of the optimal regularization 
scale are consistent with each other. Both shall be used and 
compared in the final analysis. 
Figure \ref{fig:Kehfeimrhovsmpisqnew}
indicates an optimal value 
$\hat{m}_{\pi,\ro{max}}^2 = 0.20$ GeV$^2$ for the complete data set, which 
includes 
the low-energy data.  
A higher density of data in the low-energy region serves to decrease 
the statistical error estimate of extrapolations to the low-energy region. 
The corresponding value of $\La^\ro{scale}$ is unconstrained in this case, 
since the data lie close to the PCR. 
The breakdown of the systematic error bar into its constituent uncertainties 
is listed in Table \ref{table:bd}.

\begin{table*}[tp]
  \caption{\footnotesize{Results for the quenched $\rho$ meson mass for different values of $m_{\pi,\ro{max}}^2$, extrapolated to the physical point, corresponding to Figures \ref{fig:Kehfeimrhovsmpisq} (for the original data set) and \ref{fig:Kehfeimrhovsmpisqnew} (for the complete data set). The uncertainty in $m_{\rho,Q}^{\ro{ext}}(m_{\pi,\ro{phys}}^2)$ is provided in the following order: the statistical uncertainty, the systematic uncertainty due to the intrinsic scale, $g_2$, $g_4$, $M_0^2$, $A_0$ and the regulator functional form, respectively.   
}}
\vspace{-6pt}
  \newcommand\T{\rule{0pt}{2.8ex}}
  \newcommand\B{\rule[-1.4ex]{0pt}{0pt}}
  \begin{center}
    \begin{tabular}{lll}
      \hline
      \hline
      \T\B            
      $m_{\pi,\ro{max}}^2$(GeV$^2$)  &  \qquad $m_{\rho,Q}^{\ro{ext}}(m_{\pi,\ro{phys}}^2)$ (GeV): original set &  \qquad $m_{\rho,Q}^{\ro{ext}}(m_{\pi,\ro{phys}}^2)$ (GeV): complete set\\
      \hline
      $0.059$  &\T\qquad\quad -    &\qquad$0.956(234)(1)(0)(0)(1)(0)(0)$   \\
      $0.068$  &\T\qquad\quad -   &\qquad$0.938(139)(1)(1)(0)(1)(0)(0)$   \\
      $0.080$   &\T\qquad\quad -    &\qquad$0.934(87)(1)(1)(0)(1)(0)(0)$   \\
      $0.094$  &\T\qquad\quad -   &\qquad$0.939(64)(2)(1)(0)(1)(0)(1)$   \\
      $0.114$   &\T\qquad\quad -   &\qquad$0.930(47)(3)(2)(0)(2)(0)(0)$   \\
      $0.143$  &\T\qquad\quad -   &\qquad$0.932(34)(5)(3)(0)(4)(1)(0)$   \\
      $0.172$  &\T\qquad\quad -  &\qquad$0.929(31)(6)(4)(0)(5)(1)(0)$   \\
       $0.204$  &\T\qquad\quad -  &\qquad$0.929(29)(9)(5)(0)(6)(1)(0)$   \\
      $0.241$   &\T\qquad $0.930(110)(27)(14)(0)(17)(4)(6)$ &\qquad$0.927(27)(12)(7)(0)(9)(2)(0)$   \\
        $0.288$ &\T\qquad$0.899(62)(31)(1)(0)(1)(1)$ &\qquad$0.933(24)(17)(10)(0)(12)(3)(1)$   \\
        $0.347$ &\T\qquad $0.922(43)(37)(11)(0)(13)(28)(17)$ &\qquad$0.932(23)(21)(11)(0)(13)(3)(4)$ \\
        $0.422$   &\T\qquad $0.948(29)(45)(23)(1)(28)(7)(8)$ &\qquad$0.930(20)(29)(14)(0)(16)(4)(8)$  \\
        $0.515$   &\T\qquad $0.932(23)(51)(19)(1)(23)(6)(19)$ &\qquad$0.929(19)(37)(15)(0)(18)(4)(13)$ \\
        $0.610$   &\T\qquad $0.921(18)(63)(18)(1)(22)(5)(25)$ &\qquad$0.926(16)(54)(18)(0)(21)(5)(22)$   \\
        $0.676$   &\T\qquad $0.915(12)(74)(18)(1)(22)(6)(32)$ &\qquad$0.926(14)(66)(19)(1)(23)(6)(25)$   \\
      $0.743$   &\T\qquad $0.919(13)(79)(22)(1)(26)(7)(29)$  &\qquad$0.922(12)(74)(21)(1)(26)(7)(27)$   \\
      $0.867$   &\T \qquad $0.933(9)(103)(32)(1)(39)(12)(35)$ &\qquad$0.923(8)(100)(29)(1)(35)(11)(38)$ \\ 
      $1.062$   &\T   \qquad $0.928(7)(115)(32)(1)(38)(12)(43)$  &\qquad\quad - \\
      $1.212$   &\T   \qquad $0.926(7)(121)(31)(1)(37)(12)(38)$  &\qquad\quad - \\
      $1.329$   &\T   \qquad $0.921(6)(132)(31)(1)(37)(12)(43)$ &\qquad\quad -  \\
      $1.742$   &\T   \qquad $0.915(5)(146)(30)(1)(37)(13)(48)$ &\qquad\quad -  \\
      $2.187$   &\T   \qquad $0.910(4)(175)(30)(1)(37)(14)(55)$ &\qquad\quad -  \\
      $3.150$  &\T   \qquad $0.902(3)(197)(29)(1)(36)(14)(61)$  &\qquad\quad - \\

      \hline
    \end{tabular}    
  \end{center}
  \label{table:bd}
\end{table*}

\begin{table*}[tp]
 \caption{\footnotesize{The values of $c_0$, $c_2$ and $c_4$ as obtained from both the original data set and the complete set, which includes the low-energy data. In each case, the coefficients are evaluated using the scale $\La_{\ro{central}}$ (for a triple-dipole regulator) as obtained from the $\chi^2_{dof}$ analysis. The value of $m_{\pi,\ro{max}}^2$ used is that which yields the smallest error bar in adding statistical and systematic uncertainties in quadrature. For the initial data set,  $\hat{m}_{\pi,\ro{max}}^2 = 0.35$ GeV$^2$. For the complete data set, $\hat{m}_{\pi,\ro{max}}^2 = 0.20$ GeV$^2$. The statistical uncertainty is quoted in the first pair of parentheses, and the systematic uncertainty is quoted due to the parameters, in the following order: $\Lambda^{\ro{scale}}$, $g_2$, $g_4$, $M_0^2$, $A_0$ and the regulator functional form. }}
  \newcommand\T{\rule{0pt}{2.8ex}}
  \newcommand\B{\rule[-1.4ex]{0pt}{0pt}}
  \begin{center}
    \begin{tabular}{llll}
      \hline
      \hline
       \T\B 
       \quad & $c_0$(GeV$^2$)  & \quad$c_2$ & \,\,$c_4$(GeV$^{-2}$)   \\
      \hline
%



      original set  &\T $1.31(5)(10)(4)(0)(5)(4)(8)$ & \quad$7.9(4)(25)(2)(0)(2)(1)(4)$ & \,\,$-16.2(7)(382)(3)(0)(3)(1)(4)$ \\
      complete set   &\T $1.35(4)(3)(36)(16)(60)(166)(113)$ & \quad$6.8(5)(17)(13)(1)(17)(14)(11)$ & \,\,$-3.3(16)(359)(23)(1)(28)(12)(1)$ \\

      \hline
    \end{tabular}
  \end{center}
\vspace{-6pt}
  \label{table:cscomp}
\end{table*}

The values of $c_0$, $c_2$ and $c_4$ for both the original data set 
and the complete data set are shown in Table \ref{table:cscomp}, 
with statistical 
error estimate quoted first and systematic uncertainty due to the parameters 
$\La^\ro{scale}$, $g_2$, $g_4$, $M_0^2$, $A_0$, and the regulator functional
 form quoted second, in this order. 
 In the case of the original data set, the value of 
$c_4$ is not well-determined, due to the small number of data points used. 
In the case of the complete data set, the results are dominated by 
statistical uncertainty and also results in an almost 
unconstrained value of $c_4$. Even if $\Lambda^{\ro{scale}}$ is quite 
well-determined, as observed in Figures \ref{fig:Kehfeichisqdofc0} through 
\ref{fig:Kehfeichisqdofc4}, the value of $c_4$ itself is sensitive to 
changes in the regularization scale $\Lambda$, as evident from 
Figure \ref{fig:Kehfeic4}.  
The coefficients of the complete set are less 
well-determined due to the fact that $\hat{m}_{\pi,\ro{max}}^2 = 0.20$ GeV$^2$, 
leaving only low-energy results with large statistical uncertainties 
for fitting.

\begin{figure}[tp]
\begin{center}
%
\includegraphics[height=0.80\hsize,angle=90]{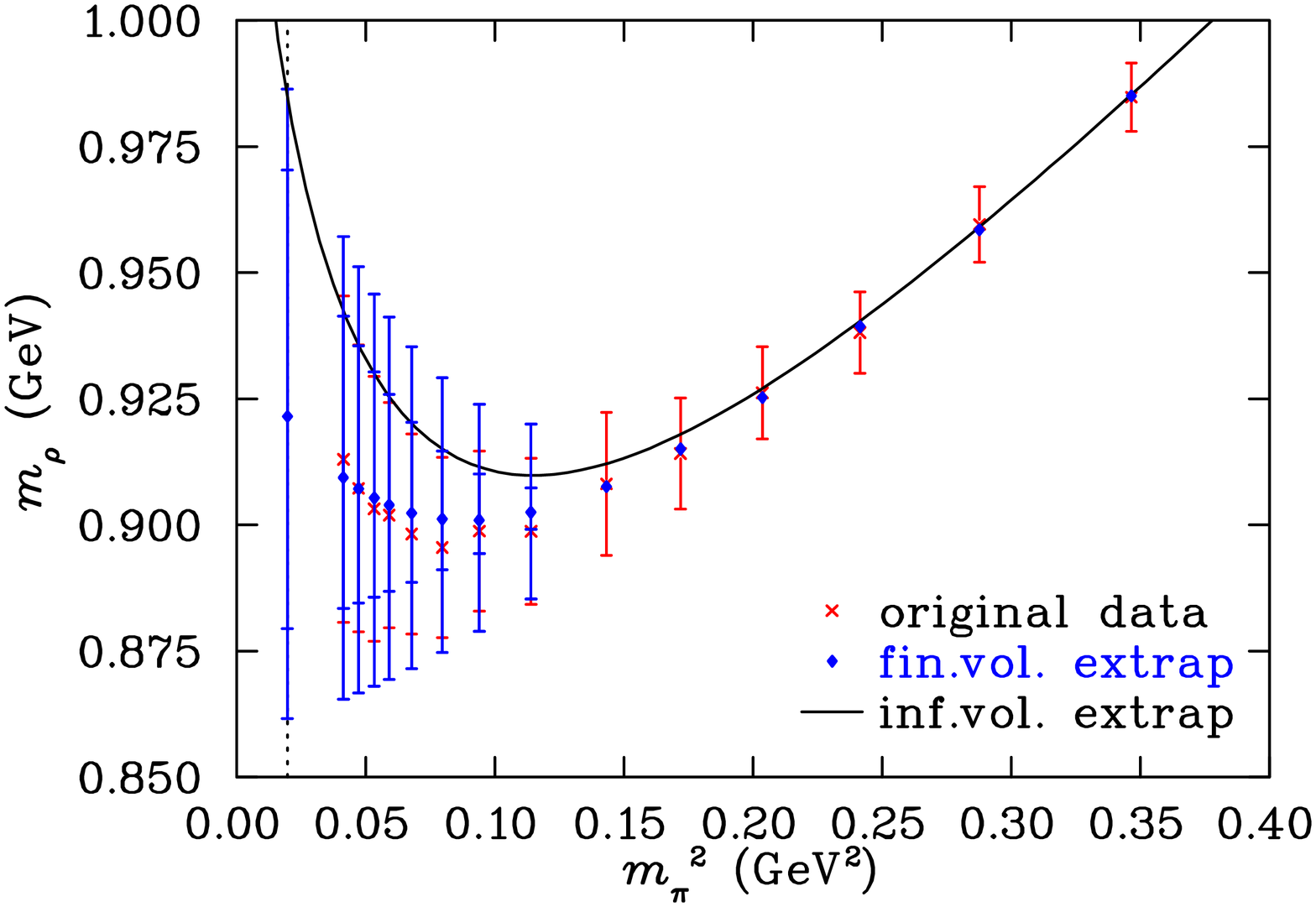}
\vspace{-11pt}
\caption{\footnotesize{(color online).  Comparison of chiral extrapolation predictions (blue diamond) with Kentucky Group data (red cross).  Extrapolation is performed at $\La_{\ro{scale}} = 0.64$ GeV, varied across the whole range of $\Lambda$ values, and using the optimal number of data points, corresponding to $\hat{m}_{\pi,\ro{max}}^2 = 0.35$ GeV$^2$. The inner error bar on the extrapolation points represents purely the systematic error from parameters. The outer error bar represents the systematic and statistical error estimates added in quadrature.}}
\label{fig:finalextrapvar}
\vspace{5pt}
\includegraphics[height=0.80\hsize,angle=90]{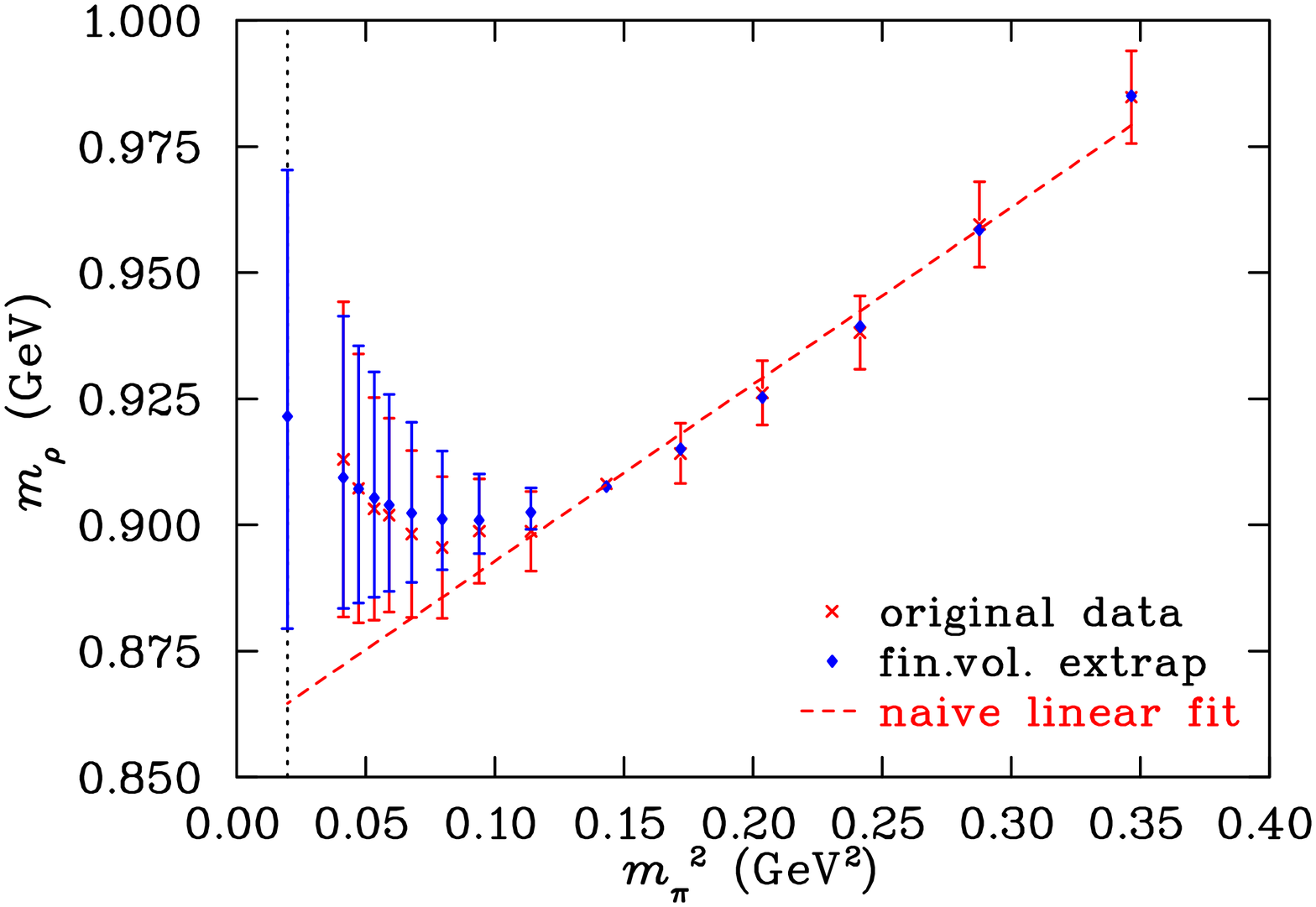}
\vspace{-11pt}
\caption{\footnotesize{(color online).  Comparison of chiral extrapolation predictions (blue diamond) with Kentucky Group data (red cross), with errors correlated relative to the point at $m_\pi^2 = 0.143$ GeV$^2$.  This is done simply to clarify the plot in Figure \ref{fig:finalextrapvar} by removing much of the correlated statistical error. Extrapolation is performed at $\La_{\ro{scale}} = 0.64$ GeV, varied across the whole range of $\Lambda$ values, and using the optimal number of data points, corresponding to $\hat{m}_{\pi,\ro{max}}^2 = 0.35$ GeV$^2$. The error bar on the extrapolation points represents the systematic error only. A simple linear fit, on the optimal pion mass region, is included for comparison.}}
\label{fig:finaldeltaextrapvar}
\end{center}
\end{figure}

The result using the estimate of the optimal regularization scale 
$\La^\ro{scale} = 0.64$ GeV, with the systematic uncertainty 
calculated by varying $\Lambda$ across all suitable values, 
and using the initial data set, is shown in Figure \ref{fig:finalextrapvar}.
The extrapolation to the physical point obtained for this quenched data set is: 
$m_{\rho,Q}^{\ro{ext}}(m_{\pi,\ro{phys}}^2) = 0.922^{+0.065}_{-0.060}$ GeV,  
an uncertainty 
of approximately $7$\%. 
Figure \ref{fig:finaldeltaextrapvar} shows the data plotted with error bars 
correlated relative to the lightest data point in the original set, 
$m_\pi^2 = 0.143$ GeV$^2$, using 
$\La_{\ro{scale}} = 0.64$ GeV, and varying $\Lambda$ across its full range 
of values. This naturally increases the estimate of the systematic 
uncertainty of the extrapolations, but also serves to demonstrate 
how closely the results from lattice QCD and $\chi$EFT match.

\section{Conclusion}
\label{sect:conc}

%

A technique for isolating an optimal regularization scale, established in 
Ref.~\cite{Hall:2010ai}, 
was tested in 
quenched QCD through an examination of the quenched $\rho$ meson mass. 
The result is a successful extrapolation
 based on an extended effective field theory procedure. 
By using 
quenched lattice QCD results that extended beyond the power-counting regime, 
an optimal regularization scale was 
obtained from the renormalization flow of the 
low-energy coefficients.  
 
An optimal value of the maximum pion 
mass to be used for fitting was also calculated, and this resulted 
in an alternative estimate of the value of the optimal regularization scale, 
which was consistent with the first result. 
%
 The mass of the $\rho$ meson was calculated in the low-energy region, 
including the physical point, using each estimate of the optimal regularization 
scale, and both results were compared.
The results of extrapolations using $\chi$EFT, and 
the results of lattice QCD simulations, were demonstrated to be consistent. 
The extrapolation correctly predicts the low-energy curvature 
that was observed when the low-energy lattice simulation results were revealed. 
%
%


In full QCD, using dynamical fermions, the process $\rho\rightarrow\pi\pi$ 
contributes to the $\rho$ meson mass. This means that 
 near the chiral limit, the $\pi\pi$ component of the $\rho$ necessarily 
involves a hard momentum scale, 
and therefore 
is not amenable to the standard methods of low-energy expansions, as entailed 
by $\chi$PT. Therefore, one needs to resort to alternative techniques in such 
instances. 

However, since there exists no experimental value for the mass of a particle in 
the quenched approximation, 
this analysis demonstrates the ability of the technique 
 to make predictions without phenomenologically motivated bias. 
The results clearly indicate a successful procedure for using  
lattice QCD data outside the power-counting regime 
to extrapolate an observable to the chiral regime.

\begin{acknowledgments}

We would like the thank Professor 
T. Cohen for helpful discussions. This research 
is supported by the Australian Research Council through Grant DP110101265. 
 Thanks go to U.S. DOE Grant No. DE-FG05-84ER40154 for partial support.  
 The research of N. Mathur 
is supported under grant No. DST-SR/S2/RJN-19/2007, India. 
The research of 
J. B. Zhang is supported by 
Chinese NSFC Grant No. 10835002 and Science Foundation of Chinese University. 
\end{acknowledgments}

\bibliographystyle{apsrev}
\bibliography{qrhoref}


\end{document}

%% file: nc.tex
\newcommand{\nc}{\newcommand}
\nc{\lb}{\langle}
\nc{\rk}{\rangle}
\nc{\Blb}{\Big\langle}
\nc{\Brk}{\Big\rangle}

\nc{\mi}{\!\!\mid\!\!}
\nc{\ra}{\rightarrow}
\nc{\Ra}{\Rightarrow}
\nc {\cd}{\partial}
\nc {\sla}{\slashed}
\nc{\ro}{\mathrm}
\nc{\ca}{\mathcal}
\nc{\bo}{\mathbf}
\nc{\Tr}{\ro{Tr}\,}
\nc{\Str}{\ro{Str}}
\nc{\realtrace}{\ro{Re\; Tr}}
\nc{\maxrealtrace}{\ro{max\, Re\; Tr}}
\nc{\ud}{\ro{d}}
\nc{\nn}{\nonumber}

\nc{\pb}{\bar{\psi}}
\nc{\p}{\psi}
\nc{\Pb}{\bar{\Psi}}
\nc{\vp}{\vec{\pi}}
\nc{\vap}{\varphi}
\nc{\vt}{\vec{\tau}}
\nc{\si}{\sigma}
\nc{\Si}{\Sigma}
\nc{\g}{\gamma}
\nc{\G}{\Gamma}
\nc{\la}{\lambda}
\nc{\La}{\Lambda}
\nc{\ep}{\epsilon}
\nc{\de}{\delta}
\nc{\De}{\Delta}
\nc{\cL}{\ca{L}}
\nc{\cLe}{\ca{L}_{\ro{eff}}}
\nc {\ti}{\tilde}
\nc{\f}{\frac}
\nc{\da}{\dagger}
\nc{\SU}{\ro{SU}}
\nc{\om}{\omega}
\nc{\Om}{\Omega}

\nc{\darrow}{\stackrel{\leftrightarrow}{\cd}}
\nc{\darrows}{\stackrel{\leftrightarrow}{\sla{\cd}}}
\nc{\Darrows}{\stackrel{\leftrightarrow}{\sla{D}}}
\nc{\mr}{\stackrel{\circ}{m}_\rho}

\nc {\eqb}{\begin{equation}}
\nc {\eqe}{\end{equation}}
\nc {\eqab}{\begin{eqnarray}}
\nc {\eqae}{\end{eqnarray}}